# Minkowski-Space Modeling of Hyperbolic Lenses

Enrico Maria Renzi[1,2], Simon Yves[1], Sveinung Erland[3], Diana Strickland[1,2], Eitan Bachmat[4], Andrea Alù[1,2]

[1]*Photonics Initiative, Advanced Science Research Center, City University of New York, New York, NY,10031, USA*
[2]*Physics Program, The Graduate Center, City University of New York, NY, 10026, USA*
[3]*Department of Mechanical Engineering and Maritime Studies, Western Norway University of Applied Sciences, 5528 Haugesund, Norway*
[4]*Department of Computer Science, Ben-Gurion University of the Negev, Beer Sheva, 84105, Israel*

*aalu@gc.cuny.edu*

The extreme anisotropy of hyperbolic materials enables extreme wave confinement, but it is also associated with an inherent misalignment between phase and energy flow, which complicates device modeling and design. Here we introduce a Minkowski-space approach to describe hyperbolic wave propagation, showing that this complexity is geometric rather than physical. By embedding anisotropy into an effective Lorentzian metric, we establish a rational design framework for hyperbolic interfaces and lenses, and analytically derive their transfer function and resolution limits, enabling ultra-large numerical apertures and deep sub-diffraction focusing. We validate our theory with the design and full-wave modeling of a planar van der Waals polaritonic lens operating in the mid-infrared frequency range.

The design of interfaces that precisely control energy-flow trajectories is central across physics and engineering, from optical lenses and microwave antennas to phononic wave concentrators. Yet, isotropic materials with circular iso-frequency dispersion [Fig. 1a] are constrained by Abbe's diffraction limit and index mismatch, fundamentally limiting wave confinement and performance [1,2]. Hyperbolic media have been proposed to overcome these limits across several wave domains, including phononics [3,4] and photonics [5-11]. Their extreme anisotropy, based on which two principal components of the constitutive tensor differ in sign, supports open (hyperbolic) dispersion contours [Fig. 1c] enabling wave propagation with exceptionally large momenta and directionality [6-12]. Despite these advantages, a longstanding challenge has been hindering the systematic design of photonic and phononic components based on these materials: phase wavefronts and energy flow trajectories are non-orthogonal, which prevents the use of ray optics and conventional geometric design rules [12]. This difficulty is particularly acute in van der Waals crystals, which naturally support hyperbolic phonon polaritons—light-matter quasiparticles arising from strong photon-phonon coupling [13-19]. There, extreme anisotropy and deeply subwavelength confinement yield strongly tilted energy-flow trajectories that defy Euclidean ray intuition, despite their ray-like appearance.

In this Letter, we show that this hurdle is geometric rather than physical: extreme anisotropy induces an effective Lorentzian metric in these materials, such that hyperbolic waves propagate in a Minkowski-type space [20] emerging directly from the constitutive relations. This mapping—associating one spatial axis of the hyperbolic medium with a temporal coordinate [21,22]—has been used to emulate cosmological models in hyperbolic metamaterials [23-27] and to optimize complex unidirectional flow processes [28, 29]. Here, we leverage this mapping in a convenient way by embedding the material anisotropy into an effective spacetime metric. This operation in turn restores orthogonality between wavefronts and rays by using generalized light-cone coordinates [30].

This geometric framework enables a rational ray-physics-based strategy to design devices leveraging boundaries between distinct hyperbolic media. In Minkowski space, classical Descartes lens constructions, developed for isotropic media, extend naturally onto hyperbolic materials. Beyond practicality, this framework reveals key physical insights: the opposite curvature of hyperbolic lenses arises from the inversion of Fermat principle, with rays following geodesics of maximal rather than minimal action. We derive transfer function and resolution limits of hyperbolic lenses, showing that, when dissipation and nonlocality are negligible, such lenses can achieve ultra-large numerical aperture (NA) and ultra-small resolution. We validate the generality of our approach with a planar hyperbolic lens in a α-$MoO_3$-$SiO_2$-gold heterostructure operating at mid-infrared frequencies. Here the lens profile follows directly from the Minkowski construction, substantially simplifying both three-dimensional and planar designs compared to full electromagnetic modeling of surface-wave focusing [31-36]. Overall, we establish a unified geometric foundation for understanding and designing lenses, concentrators and other devices in hyperbolic materials, spanning across different frequency regimes, material platforms and wave domains, and holding the promise of enhanced performance metrics compared to isotropic or weakly anisotropic materials.

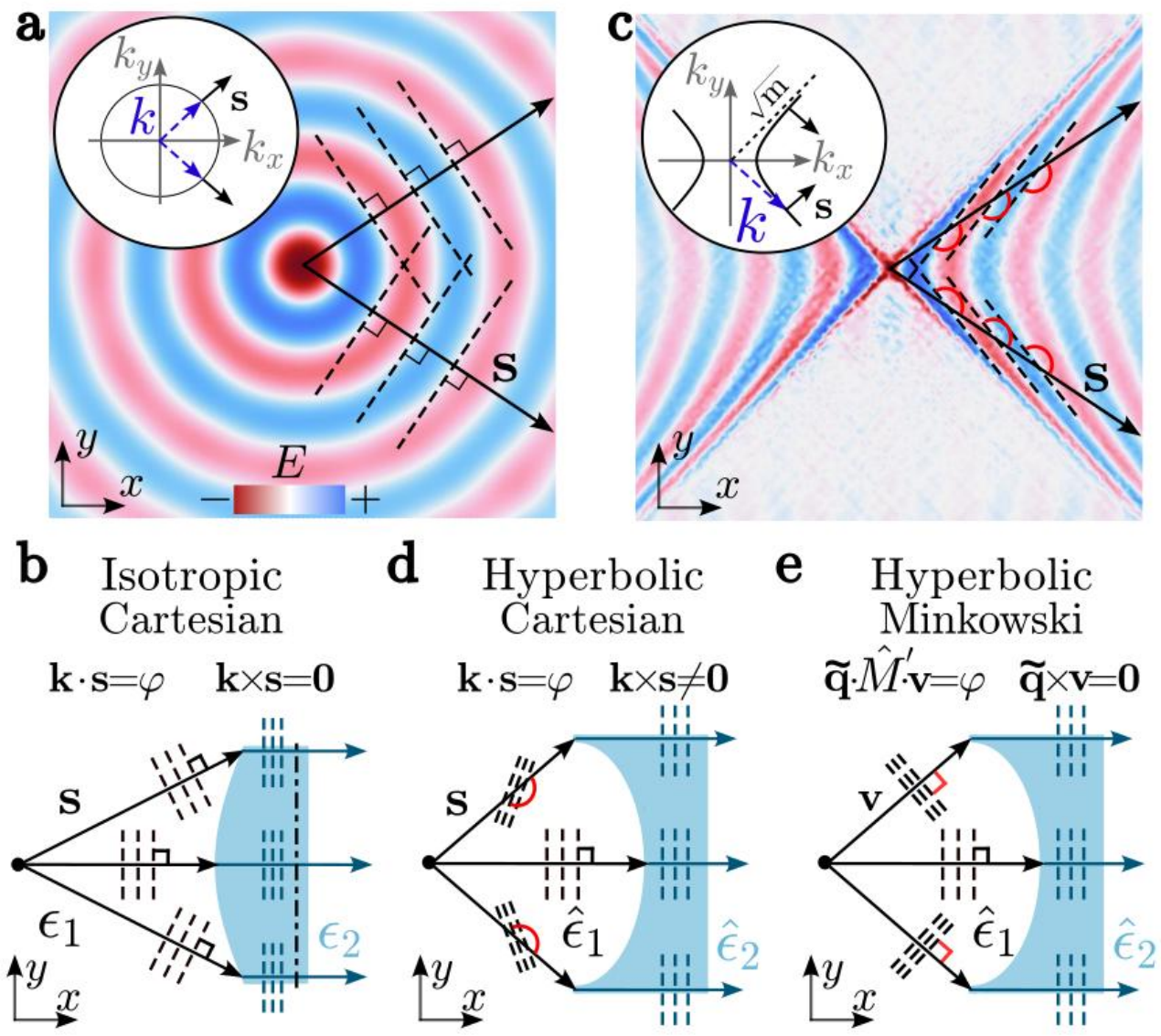

**Fig. 1. Wave propagation and lens design in hyperbolic media: Cartesian vs. Minkowski geometry.** (a) In isotropic media, wave propagation features a circular dispersion (inset). This ensures that momentum **k** and ray direction **s** are parallel, and wavefronts (dashed) orthogonal to rays (arrows), i.e., $\mathbf{k}\times\mathbf{s}=0$. b) This facilitates lens (light blue) design between two isotropic media with permittivity $\epsilon_1$ and $\epsilon_2$, following conventional ray-optics principles. (c) In hyperbolic media, dispersion contours are hyperbolic (inset), causing misalignment between **k** and **s** (angles indicated in red), i.e., $\mathbf{k}\times\mathbf{s}\neq 0$. (d) In Cartesian coordinates, constructing lenses between hyperbolic media with permittivity tensors $\hat{\epsilon}_1$ and $\hat{\epsilon}_2$, is therefore challenging. However, (e) working in a Minkowski metric $\hat{M}$ such that $\varphi=\tilde{\mathbf{q}}\cdot\hat{M}'\mathbf{v}$, where $\tilde{\mathbf{q}}$ and $\mathbf{v}$ are the group velocity and ray displacement, respectively, restores $\tilde{\mathbf{q}}\times\mathbf{v}=0$, enabling the use of ray-optics design rules for hyperbolic media.

Wave propagation in isotropic materials follows a circular iso-frequency dispersion [Fig. 1a, inset], fully described by Euclidian geometry. In this case [Fig. 1a], energy flow **s** and phase fronts are aligned with the wave-vector **k**. Lens design reduces to geometric phase-compensation: to form an image, rays from a point source must accumulate identical optical phase $\varphi=\mathbf{k}\cdot\mathbf{s}$ (or time delay) at the image plane [Fig. 1b, black dot]. This underlies Descartes optics, but is limited by the lens finite numerical aperture and by the evanescent decay of waves with transverse momentum $k_y > k_0\sqrt{\epsilon_1}$ ($k_0=\omega c_0^{-1}$), which do not contribute to imaging and lead to Abbe's resolution limit [1].

In stark contrast, hyperbolic materials depart from this paradigm. Their dispersion [Fig. 1c, inset] allows arbitrarily large wavevectors in the absence of loss and nonlocality. Consider a linear hyperbolic material with permittivity tensor $\hat{\epsilon} = \mathrm{diag}(-\epsilon_{xx}, \epsilon_{yy}, \epsilon_{yy})$. For propagation in the $xy$ plane, the dispersion admits real wavevectors satisfying $F(k_x, k_y) = k_x^2/\epsilon_{yy} - k_y^2/\epsilon_{xx} - k_0^2 = 0$ [Fig. 1c, inset], $\epsilon_{xx}\epsilon_{yy} > 0$. In principle, this allows arbitrarily large wavevectors, up to $k_y = \sqrt{m}k_x$ ($m = \epsilon_{yy}/\epsilon_{xx}$), suggesting unbounded confinement and resolution. However, extreme anisotropy breaks geometric optics: wavevector and energy flow are non-collinear ($\mathbf{k}\times\mathbf{s} \neq 0$) [Fig. 1c,d], undermining conventional ray design. Thus hyperbolic lenses [Fig. 1d] typically rely on full-wave analysis or brute-force numerical optimization [1,31-36].

Remarkably, wave propagation in hyperbolic media is naturally described by a Lorentz metric [20], requiring a Minkowski-like geometry where one spatial coordinate plays the role of time [21]. We therefore reformulate ray optics in hyperbolic media using a Minkowski-space formalism. In this setting, the apparent failure of ray optics is geometric rather than physical. Within this framework, wavevectors and energy flow recover the isotropic geometric relationship. Thus, the Euclidean non-orthogonality is a coordinate artifact of an inappropriate metric, not a fundamental property of hyperbolic waves. This description becomes exact when wave propagation preserves polarization separability, which occurs in uniaxial media and in biaxial media restricted to principal planes, where the Fresnel surface factorizes into independent isotropic and hyperbolic branches governed by Euclidean and Minkowski geometries, respectively [37].

Fundamentally, ray propagation in hyperbolic media reduces to defining distance and evaluating phase accumulation. Phase accumulation is always the scalar product of momentum and displacement. In isotropic media, this reduces to the familiar Euclidean dot product $\varphi = \mathbf{k}\cdot\mathbf{s}$, since the underlying metric is the identity, and wavevector and energy flow are covariant, while distance is $\Delta s = \sqrt{\Delta x^2 + \Delta y^2}$. In hyperbolic media, the same scalar-product remains, but Euclidean distance fails. In fact, the correct metric is $\hat{M} = \epsilon_{yy}\hat{\mathbf{x}}\hat{\mathbf{x}} - \epsilon_{xx}\hat{\mathbf{y}}\hat{\mathbf{y}}$ corresponding to a (1+1)-Minkowski spacetime [21], with Lorentzian distance $\Delta s = \sqrt{\epsilon_{yy}\Delta x^2 - \epsilon_{xx}\Delta y^2}$ [37]. Although this metric captures the physics, phase evaluation in Cartesian coordinates is cumbersome because wavevector and energy flow are non-parallel. The key simplification comes from generalized light-cone coordinates [30] adapted to the anisotropy.

To illustrate this feat, consider an interface between two unbounded hyperbolic media, $\hat{\epsilon}_1$ and $\hat{\epsilon}_2$. The iso-frequency contours [Fig. 2a] share the same asymptotic apertures $m = \epsilon_{j,xx}/\epsilon_{j,yy}$ and vertices $\left(k_0\sqrt{\epsilon_{j,yy}}, 0\right)$, $\epsilon_{2,yy} > \epsilon_{1,yy}$, so $j$=1 and $j$=2 are the optically fast and slow media, respectively. The appropriate stretched and rotated coordinate are $(x', y') = 2^{-1/2}\left(x - \sqrt{m}y, x + \sqrt{m}y\right)$ [Fig. 2b, top]. In these coordinates, the Minkowski metric becomes $\hat{M}' = \hat{\mathbf{x}}\hat{\mathbf{y}} + \hat{\mathbf{y}}\hat{\mathbf{x}}$ and the dispersion equation reduces to $q_{x'}q_{y'} = k_0^2\epsilon_{yy}/2$, rendering the medium effectively isotropic [37]. Phase accumulation reads, independently of the material aperture $m$ [Fig. 2b, bottom],

$$\varphi = \tilde{\mathbf{q}}\cdot\hat{M}'\mathbf{v} = k_0\sqrt{\epsilon_{j,yy}}\sqrt{2\Delta x'\Delta y'}. \quad j = 1, 2. \tag{1}$$

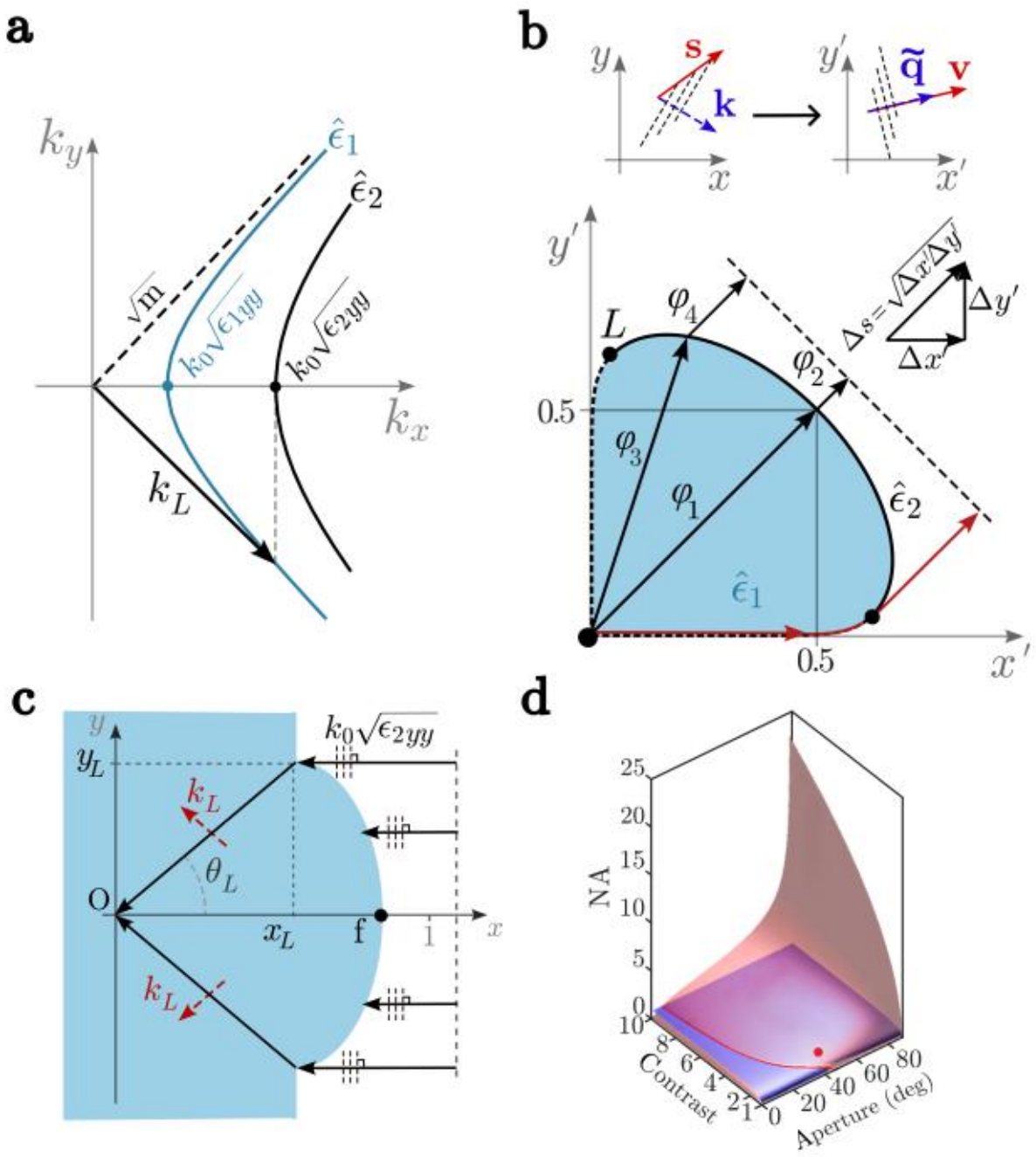

**Fig. 2. Lens equation and numerical aperture.** (a) Iso-frequency contours of two hyperbolic media $\hat{\epsilon}_{\{1,2\}} = \mathrm{diag}(\epsilon_{\{1,2\},xx}, \epsilon_{\{1,2\},yy})$ with matched apertures $\sqrt{m}$, vertices $k_0\sqrt{\epsilon_{\{1,2\},yy}}$ and refractive index contrast $T = \sqrt{\epsilon_{2,yy} / \epsilon_{1,yy}}$. (b) Collimator (bottom) construction in the reference frame $(x', y')$ (inset), where momentum and ray are collinear. Beyond point $L$, rays reach the interface at the total internal reflection threshold, enabling adiabatic outcoupling and preventing diffraction at the lens boundaries. (c) Schematic of a hyperbolic focusing singlet, with maximal acceptance angle set by the line of sight $(x_L, y_L)$. (d) Numerical aperture plotted as a function of contrast $T = \epsilon_{2,yy} / \epsilon_{1,yy}$ and k-space aperture angle $\tan^{-1}(\sqrt{m})$. The NA increases with both parameters and is virtually unbounded for large contrast and wide apertures. Lenses in (b) and (c) are obtained with $\hat{\epsilon}_1 = \mathrm{diag}(-1,1)$ and $\hat{\epsilon}_2 = \mathrm{diag}(-4,4)$, giving $m = 1$ and $T = 2$.

Here, $\Delta s' = \sqrt{2\Delta x' \Delta y'}$ is the generalized distance along the displacement vector $\mathbf{v} = [\Delta x', \Delta y']^T$. Crucially, in light-cone coordinates, the contravariant ray vector $\tilde{\mathbf{q}} = k_0\sqrt{\epsilon_{yy} / 2\Delta x' \Delta y'}[\Delta x', \Delta y']^T$, aligned with energy flow, directly yields the phase [37]. Here $\mathbf{q}$ is a covariant wavevector and $\tilde{\mathbf{q}}$ its contravariant counterpart. Since the metric is diagonal in light-cone coordinates, $\tilde{\mathbf{q}} = (\hat{M}')^{-1}\mathbf{q}$ is reciprocal to $\mathbf{q}$ and parallel to the physical "ray" displacement $\mathbf{v}$, ($\tilde{\mathbf{q}} \times \mathbf{v} = 0$) [Fig. 1e]. Phase accumulation follows from the covariant contraction $\varphi = \mathbf{q} \cdot \mathbf{v} = \tilde{\mathbf{q}} \cdot \hat{M}'\mathbf{v}$, restoring a Descartes ray picture where phase fronts and energy flow are orthogonal.

Using Eq. (1), we construct the collimator in Fig. 2b. The front interface (black solid line) enforces phase continuity along different optical paths equating phase along the path $\varphi_1 + \varphi_2$ to that along $\varphi_3 + \varphi_4$. This yields the converging singlet

$$y = (2m)^{-1/2} \sqrt{(T-1)\left(1-T+2\sqrt{2}xT-2(T+1)x^2\right)}, \tag{2}$$

where, analogously with isotropic media we defined the refractive index contrast as $T = \sqrt{\epsilon_{2,yy}/\epsilon_{1,yy}}$ .

Unlike its isotropic counterpart [Fig. 1b], the hyperbolic lens curvature is reversed. This inversion is not a geometric accident, but a direct consequence of the phase accumulation mechanism of hyperbolic media. In isotropic materials the lens curvature compensates for increased optical path of oblique rays, ensuring equal phase at the focus. Hyperbolic media exhibit the opposite behavior. As made explicit by Eq. (1), phase is maximal along the optical axis ($\Delta x'=\Delta y'$) and decreases monotonically with increasing transverse momentum, vanishing in the large-momentum limit ($\Delta x' \to 0$ or $\Delta y' \to 0$). To compensate for this decreasing phase of marginal rays, the lens must therefore bend in the opposite direction. As shown in [37], this reflects a deeper principle: hyperbolic rays obey action maximization rather than minimization [20], inverting Fermat's principle. On-axis rays accumulate maximal proper time and phase, and the wave behaves analogously to a massive particle at rest. In contrast, asymptotic rays acquire vanishing action, analogous to relativistic particles approaching light speed, which experience zero proper time and therefore acquire no phase. These limits provide the physical origin of curvature inversion.

Having established that hyperbolic rays maximize optical action, we next consider large-angle rays beyond $L(x_L', y_L')$ [Fig. 2b, black bullet], where the lens ceases to collimate. The remaining interface [Fig. 2b, bottom side, black dotted line], is set by requiring rays at the boundary (red arrow) to be at total-internal-reflection threshold, ensuring adiabatic outcoupling. Normally, this condition requires a rigorous application of electromagnetic boundary conditions. However, in Minkowski space, this reduces to equating the phase pick-up of a lateral ray reaching $L(x_L', y_L')$ to that of a ray launched from the origin [37], guaranteeing a seamless transition and mitigating diffraction from lateral discontinuities.

A key metric for lens performance is its numerical aperture (NA), set by the largest accepted transverse momentum. For the collimator in Eq. (2), inversion under uniform normal illumination shows that the NA is limited by the largest transverse wavevector $k_L$ [Fig. 2c]. This leads to

$$\mathrm{NA} = \sqrt{\epsilon_{1,yy}\, m(T-1)}, \tag{3}$$

Interestingly, although the geometric acceptance angle, $\theta_L$, is finite due to its limited physical aperture, the NA is virtually unbounded.

Figure 2d shows Eq. (3) versus contrast $T$ and k-space aperture $m$, for the general case $\epsilon_{1,yy} = 1$ . Unlike isotropic collimators, where $\mathrm{NA}_{iso} = \sqrt{\epsilon_1 (1-1/T)}$ and intrinsically limited, the hyperbolic NA increases with both parameters. Moreover, the hyperbolic scenario introduces a degree of freedom $m$ and, for the same index and contrast, if $m > T^{-1}$ the hyperbolic lens exceeds its isotropic counterpart [37].

To validate our Minkowski design, we apply it to hyperbolic media with $\epsilon_{1,yy} = 1$, aperture $m = 1$ and contrast $T = 2.23$ [Fig. 2d, red dot]. These parameters outperform isotropic configurations of comparable contrast and may be realized both in hyperbolic engineered materials and natural media [36]. A 1% loss term models natural dissipation and aids numerical convergence. Although no unique optimal contrast exists, we select a large aperture to maximize NA while keeping $T$ moderate to limit on-axis reflection losses below 10%.

Figure 3 illustrates two lens implementations from COMSOL Multiphysics full-wave simulations of the magnetic field $H_z$. In Fig. 3a, the lens of Fig. 2b-c focuses a normally incident paraxial beam coming from $x > 0$ to $f = -10\sqrt{2} = -12\lambda_0$. In Fig. 3b, a point dipole at $f_1 = -20\lambda_0$ is focused to the back focal plane $f_2 = 16\lambda_0$. Here the lens equation generalizes Eq. (2) to yield finite focus [37]. Maximizing the NA increases material contrast and reflection losses. These losses may be reduced by lowering contrast or adding anti-reflection coatings, warranting further investigation which is beyond the scope of this paper.

The NA describes the lens collection capability but not the resolution, which also depends on aperture, diffraction, aberrations, and coherence. We evaluate focusing by computing the Green's function in Minkowski-space and applying diffraction theory to obtain the focal plane field [37]. We consider transverse-magnetic excitation with magnetic field $H_z$ at free-space wavelength $\lambda_0$, normally incident and uniformly illuminating the aperture. The lens interface [Fig. 3a] is treated as optically thin and the focus lies in the far field ($|f| \gg y_L$, $f \ll x_L$, $|x_L| \ll \lambda_0$).

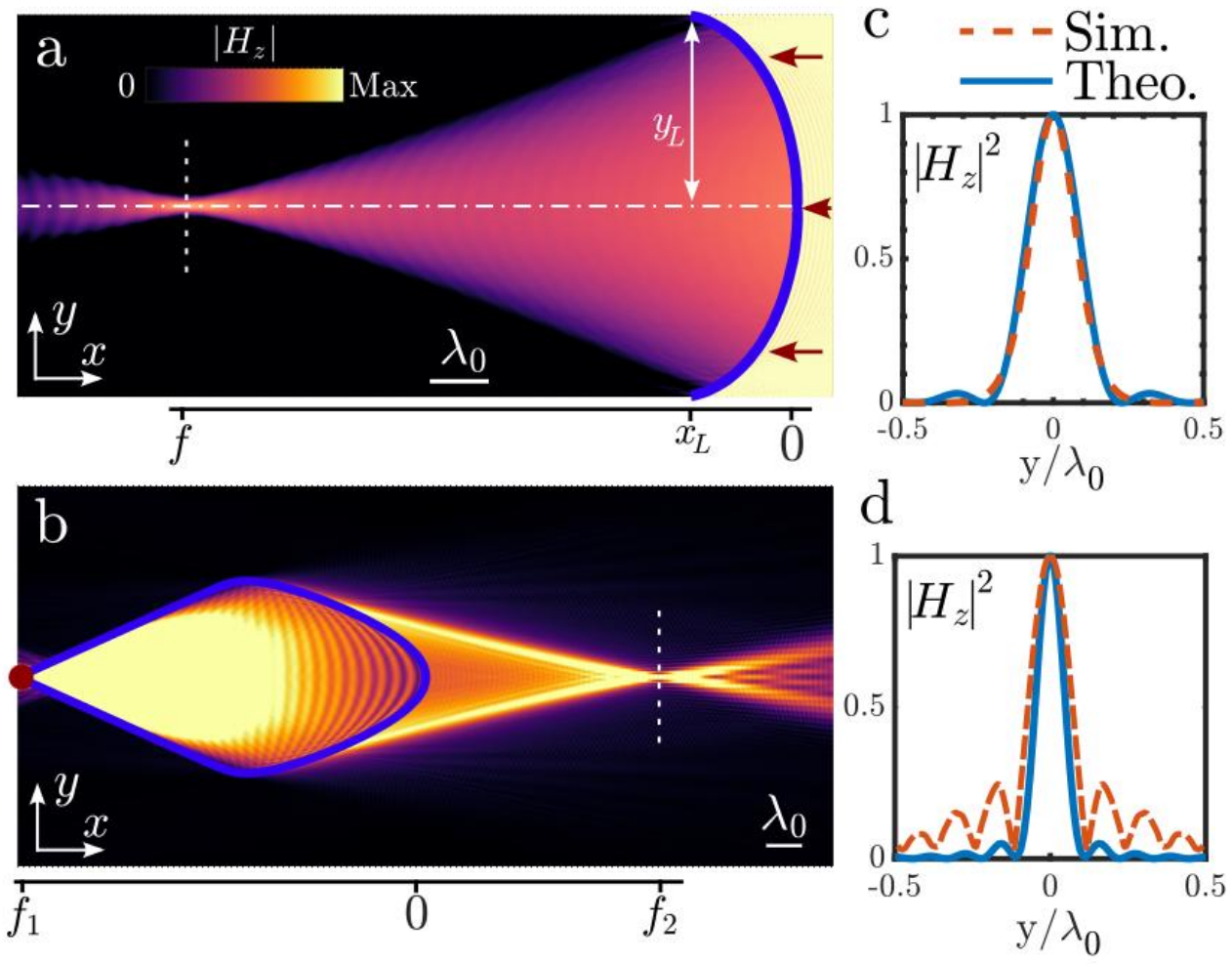

**Fig. 3. Full-wave simulations and resolution estimation**. Magnetic field distribution $|H_z|$ for (a) a hyperbolic lens focusing a paraxial beam coming from the right to a focal distance $f = -12\lambda_0$ from the lens interface, where $\lambda_0 = 2\pi / (k_0\sqrt{\epsilon_{1,yy}})$, and (b) a hyperbolic lens focusing a point dipole (red dot) at $f_1 = -20\lambda_0$ to its back focal plane at $f_2 = 16\lambda_0$. Theoretical (blue) and simulated (dashed red) point spread function for (c) the focuser and (d) for the converging singlet evaluated on their front focal plane $(f, y)$ and back focal plane $(f_2, y)$, respectively. These lenses show a resolution (a) $R = 0.44\lambda_0$ and (b) $R = 0.11\lambda_0$, respectively. The material contrast and anisotropy are $T = 2.23$ and $m = 1$.

Under these conditions, a slowly varying wavefront allows a Fresnel-type transfer function with a modified quadratic phase incorporating the material anisotropy [1]. The aperture field, with anisotropy-dependent phase, propagates to $x = f$ [Fig. 3a, white dashed line], as

$$U(f,y) \simeq \frac{1}{i\lambda_0\sqrt{2}}\sqrt{\frac{2(T-1)}{\epsilon_{1,yy}m(T+1)}}\,\mathrm{sinc}\left(\frac{2\pi}{\lambda_0}\sqrt{\frac{\epsilon_{1,yy}m(T-1)}{4(T+1)}}y\right)\exp\left[i\frac{2\pi}{\lambda_0}\sqrt{2\epsilon_{1,yy}}\left(1+\frac{m}{4}y^2\right)\right]. \tag{4}$$

The resolution is the main-lobe of Eq. (4), given by the first zero of the sinc function:

$$\mathrm{Res}{=}2\lambda_0\frac{\sqrt{4(T+1)}}{\mathrm{NA}}. \tag{5}$$

Increasing contrast improves resolution by matching larger transverse wavevectors [Fig. 2c], whereas low contrast leads to degraded performance approaching the isotropic limit. Strong in-plane anisotropy ( $m \gg 1$ ) further enhances resolution by compressing the beam, whereas weak anisotropy reduces confinement. A smaller effective index $\sqrt{\epsilon_{1,yy}}$ increases the NA and improves resolution. Although wavelength-dependent, appropriate contrast and anisotropy enhance NA to achieve subwavelength focusing beyond Abbe's limit [1]. Importantly, although Eq. (5) is unbounded, material loss ultimately limits resolution. Figures 3c-d show good agreement between simulated and theoretical focal-plane fields [Fig. 3a-b, white dashed lines] for both collimator and convex lens.

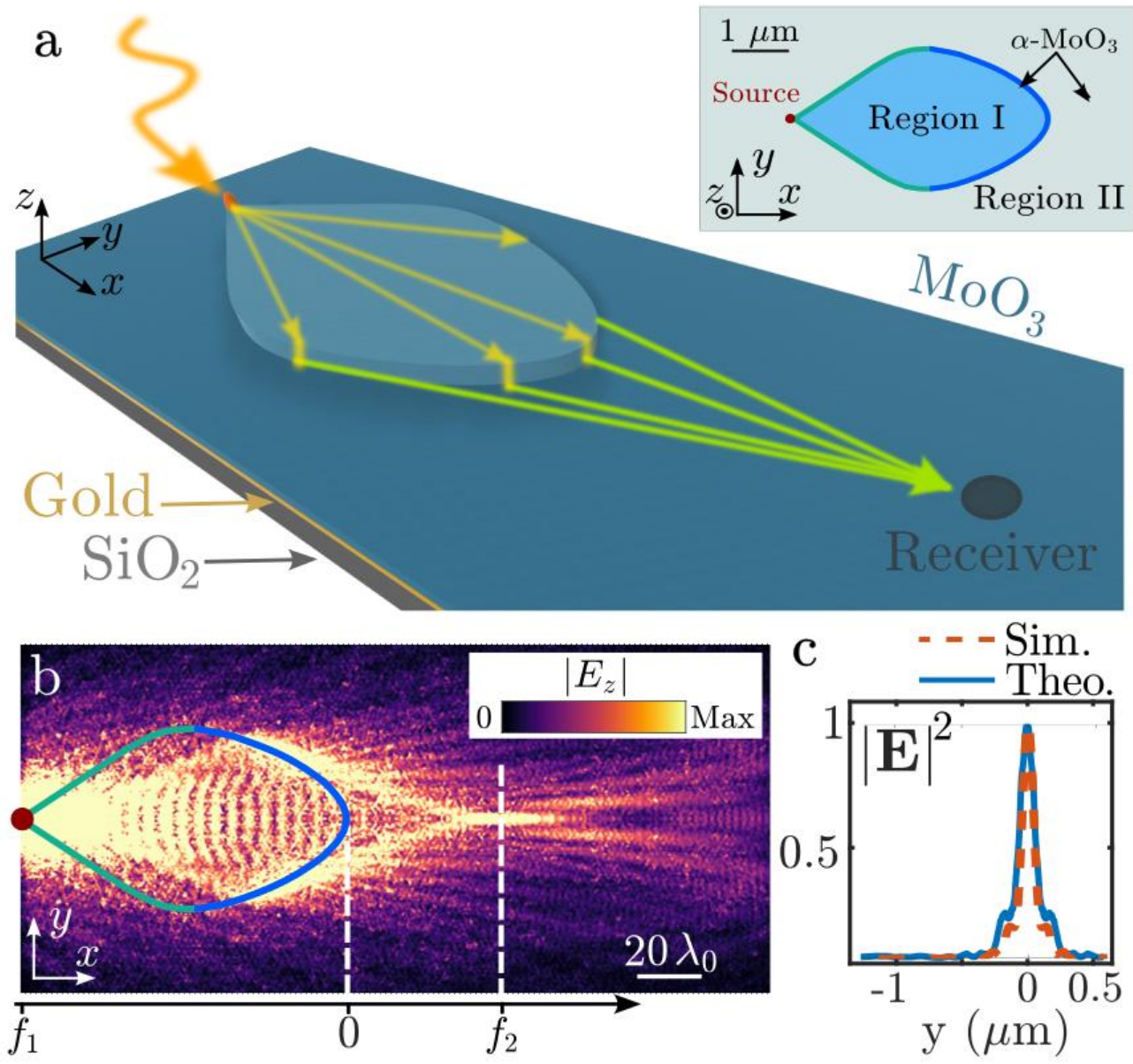

**Fig. 4. Implementation of a vdW polaritonic lens heterostructure in the mid-IR.** (a) Hyperbolic lens made from a structured α-$MoO_3$ layer on gold and $SiO_2$, focusing light from a scatterer to a receiver via hyperbolic phonon polariton propagation. The inset shows a top view of the lens geometry: the blue line denotes the collimating interface, while the green line corresponds to the welding region. Region I consists of a thicker, low-index layer, whereas Region II is a thinner, high-index region. (b) Simulated electric field distribution $|E_z|$, showing highly confined focusing. (c) Field intensity profile $|E_z|$ at the focal plane $(f_2, y)$, with a resolution $\lambda_0/42.07$ and $\lambda_p/1.9$ ( $\lambda_0 = 10.58\mu\mathrm{m}$, $\lambda_p = 480\mathrm{nm}$ ). Simulated heterostructure consisting of a 50 nm thick gold layer, $MoO_3$ layers with thickness $d_1 = 100$ nm for region I and $d_2 = 80$ nm for region II (inset), operating at 27.23 THz. Material contrast and aperture are $m_{eff} = 2.4$ and $T_{eff} = 1.73$.

To demonstrate the practicality of our framework, we design a realistic polaritonic nano-lens operating in the mid-infrared, a spectral range central to sensing and imaging applications [43-45]. Using the Minkowski strategy, we realize a compact van der Waals α-$MoO_3$ lens supporting strongly subwavelength focusing of ultra-low-loss hyperbolic surface phonon polaritons (h-SPhPs). Unlike prior polaritonic lenses relying on weak confinement or approximate profiles [33,36], this platform enables systematic implementation of theory-prescribed geometries via thickness-controlled confinement [33,36,46,47].

The structure [Fig. 4a] comprises two α-$MoO_3$ regions of different thicknesses, forming effective low- (Region I) and high-index (Region II) hyperbolic media while preserving the k-space aperture. A gold–$SiO_2$ substrate enhances confinement ensuring that the guided polariton closely approximate a hyperbolic contour over a broad momentum range [48]. From the guided-mode dispersion we extract $m_{eff} = 2.4$ and $T_{eff} = 1.73$ [37], and design a lens that focuses a point source from $f_1 = -5.65\mu\text{m}$ to $f_2 = 2.35\mu\text{m}$ at 27.3 THz [Fig. 4b]. Full-wave simulations [Fig. 4b,c] agree with theory and yield a deeply subwavelength focal spot consistent with the predicted resolution.

In this work, we have introduced a Minkowski-space formulation for wave propagation in hyperbolic media, which addresses the complexity of modeling hyperbolic media by recasting wave propagation in a geometry where phase accumulation and ray dynamics admit a unified geometric description. Within this framework, we show that the Euclidean misalignment between wavevector and energy flow is a geometric artifact, and outline a rational ray-based design of hyperbolic interfaces and lenses using principles of isotropic optics. Beyond its practical implications, the Minkowski mapping provides direct physical insight into hyperbolic wave propagation and interfaces. In the Lorentzian geometry set by the material response (expressed in light-cone coordinates), wavefronts and rays recover orthogonality and the contravariant ray vector directly evaluates phase via metric contraction, even though energy flow differs from the Euclidean phase gradient. We derived the transfer function and resolution limits of ideal hyperbolic lenses, predicting ultra-large NA and deep sub-diffractional focusing in the absence of loss and nonlocality. These predictions were validated by full-wave simulations and by a planar α-$MoO_3$ mid-IR implementation achieving subwavelength focusing of surface phonon polaritons. Overall, this geometric perspective establishes a unified framework for the rational design and physical interpretation of hyperbolic interfaces, lenses, and concentrators across natural and artificial platforms and wave systems.

The authors wish to thank Dr. Emanuele Galiffi and Dr. Emroz Khan for useful discussions and acknowledge support from the Simons Foundation and the Office of Naval Research.

## References

1. B. E. A. Saleh, M. C. Teich, "Fundamentals of Photonics," 1st ed (John Wiley & Sons, Inc., 1991).
2. J. B. Pendry, "Negative Refraction Makes a Perfect Lens," Phys. Rev. Lett. **85**, 3966 (2000).
3. J. H. Oh, H. M. Seung, and Y. Y. Kim, "A truly hyperbolic elastic metamaterial lens," *Appl. Phys. Lett.* **104** (7): 073503 (2014).
4. J. Li, L. Fok, X. Yin, G. Bartal, and X. Zhang, "Experimental demonstration of an acoustic magnifying hyperlens," *Nature Mater.* **8**, 931–934 (2009).
5. P. V. Kapitanova, P. Ginzburg, F. J. Rodríguez-Fortuno, F. S. Filonov, P. M. Voroshilov, P. A. Belov, A. N. Poddubny, Y. S. Kivshar, G. A. Wurtz, and A. V. Zayats, "Photonic spin Hall effect in hyperbolic metamaterials for polarization-controlled routing of subwavelength modes," *Nat Commun* **5**, 3226 (2014).
6. Z. Liu, H. Lee, Y. Xiong, C. Sun, and X. Zhang, "Far-Field Optical Hyperlens Magnifying Sub-Diffraction-Limited Objects," *Science* **315**, 5819 (2007).

7. Z. Jacob, L. V. Alekseyev, and E. Narimanov, "Semiclassical theory of the hyperlens," *J. Opt. Soc. Am.* A 24, A52-A59 (2007).
8. J. Rho, Z. Ye, Y. Xiong, X. Yin, Z. Liu, H. Choi, G. Bartal, and X. Zhang, "Spherical hyperlens for two-dimensional sub-diffractional imaging at visible frequencies," *Nat. Commun.* **1**, 143 (2010).
9. C. B. Ma, M. A. Escobar, Z. W. Liu, "Extraordinary light focusing and Fourier transform properties of gradient-index metalenses," *Phys. Rev. B* **84**, 195142 (2011).
10. J. Sun, M. I. Shalaev, and N. M. Litchinitser, "Experimental demonstration of a non-resonant hyperlens in the visible spectral range," *Nat. Commun.* **6**, 7201 (2015).
11. C.W. Tao, T. J.Yen, and T. Y. Huang, "Achieving sub-wavelength imaging through a flat hyperlens in a modified anodic aluminum oxide template," *Sci. Rep.* **10**, 5296 (2020).
12. A. Poddubny, I. Iorsh, P. Belov, Y. Kivshar, "Hyperbolic metamaterials," *Nature Photon* **7**, 948–957 (2013).
13. D. N. Basov , M. M. Fogler, and F. J. García de Abajo, "Polaritons in van der Waals materials," *Science* **354**, 6309 (2016).
14. E. Galiffi, G. Carini, X. Ni, G. Álvarez-Pérez, S. Yves, E. M. Renzi, R. Nolen, S. Wasserroth, M. Wolf, P. Alonso-Gonzalez, A. Paarmann, and Andrea Alù, "Extreme light confinement and control in low-symmetry phonon-polaritonic crystals," *Nat. Rev. Mat*. **9**, 9–28 (2024).
15. H. Wang, A. Kumar, S. Dai, X. Lin, Z. Jacob, S.-H. Oh, V. Menon, E. Narimanov, Y. D. Kim, J.-P. Wang, P. Avouris, L. M. Moreno, J. Caldwell, and Tony Low, "Planar hyperbolic polaritons in 2D van der Waals materials," *Nat. Commun.* **15**, 69 (2024).
16. W. Ma, P. Alonso-González, S. Li, A. Y. Nikitin, J. Yuan, J. Martín-Sánchez, J. Taboada-Gutiérrez, I. Amenabar, P. Li, S. Vélez, *et al.*, "In-plane anisotropic and ultra-low-loss polaritons in a natural van der Waals crystal," *Nature* **562**, 557–562 (2018).
17. I-H. Lee, M. He, X. Zhang, Y. Luo, S. Liu, J. H. Edgar, K. Wang, P. Avouris, T. Low, J. D. Caldwell and S-H. Oh, "Image polaritons in boron nitride for extreme polariton confinement with low losses," *Nat. Commun.* **11**, 3649 (2020).
18. G. Ni, A. S. McLeod, Z. Sun, J. R. Matson, C. Fan Bowen Lo, D. A. Rhodes, F. L. Ruta, S. L. Moore, R. A. Vitalone, R. Cusco, et al., "Long-Lived Phonon Polaritons in Hyperbolic Materials," *Nano Letters* **21** (13), 5767-5773 (2021).
19. X. Ni, G. Carini, W. Ma, E. M. Renzi, E. Galiffi, S. Wasserroth, M. Wolf, P. Li, A. Paarmann, and Andrea Alù "Observation of directional leaky polaritons at anisotropic crystal interfaces," *Nat. Commun.***14**, 1, 2845 (2023).
20. R. Penrose, *Techniques of Differential Topology in Relativity* (Society for Industrial and Applied Mathematics, J.W. Arrowsmith, Ltd., Bristol, England, 1972).
21. I. I. Smolyaninov, and E. E. Narimanov, "Metric Signature Transitions in Optical Metamaterials," *Phys. Rev. Lett.* **105**, 067402 (2010).
22. I. I Smolyaninov, "Modeling of causality with metamaterials," J. Opt. **15** 025101 (2013).
23. I. I. Smolyaninov, E. Hwang, and E. Narimanov, "Hyperbolic metamaterial interfaces: Hawking radiation from Rindler horizons and spacetime signature transitions," *Phys. Rev. B* **85**, 235122 (2012).
24. D. Figueiredo, F. Morales, S. Fumeron, and B. Berche, "Cosmology in the laboratory: An analogy between hyperbolic metamaterials and the Milne universe," *Phys. Rev. D* **96**, 105012 (2017).
25. S. H. Hendi, Z. S. Taghadomi, A. G. Ardakani, "Metamaterials mimic the black holes: the effects of charge and rotation on the optical properties," Progress of Theoretical and Experimental Physics **7**, 073E01 (2020).
26. B. Tekin, "Hyperbolic metamaterials and massive Klein-Gordon equation in (2+1)-dimensional de Sitter spacetime," *Phys. Rev*. D **104**, 105004 (2021).
27. J.-Z. She, S. Tao, T. Liu, and H. Chen, "Simulation of the expanding universe in hyperbolic metamaterials," *Opt. Express* **31**, 33312-33319 (2023).
28. S. Erland, E. Bachmat, and A. Steiner, "Let the fast passengers wait: Boarding an airplane takes shorter time when passengers with the most bin luggage enter first," European Journal of Operational Research **317**, 3 (2024).
29. E. Bachmat, S. Erland, V. Frette, and J. Kaupužs, "Lenses in Minkowski space optimize queues by hiding bottleneck units," preprint, arXiv:2306.08307v1 (2023).
30. P. A. M. Dirac, "Forms of Relativistic Dynamics," Reviews of Modern Physics **21** (392): 392–399 (1949).
31. J. Bénédicto, E. Centeno, and A. Moreau, "Lens equation for flat lenses made with hyperbolic metamaterials," *Opt. Lett.* **37**, 4786-4788 (2012).
32. J. Martín-Sánchez et al. "Focusing of in-plane hyperbolic polaritons in van der Waals crystals with tailored infrared nanoantennas," *Sci. Adv.* **7**, eabj0127 (2021).
33. J. Duan, G. Álvarez-Pérez, A.I.F. Tresguerres-Mata, et al. "Planar refraction and lensing of highly confined polaritons in anisotropic media," *Nat. Commun*. **12**, 4325 (2021).
34. S. Dehdashti, A. Shahsafi, B. Zheng, L. Shen, Z. Wang, R. Zhu, H. Chen, and H. Chen, "Conformal hyperbolic optics," *Phys. Rev. Research* **3**, 033281 (2021).
35. S. Tao, T. Hou, Y. Zeng, G. Hu, Z. Ge, J. Liao, S. Zhu, T. Zhang, C.-W. Qiu, and H. Chen, "Anisotropic Fermat's principle for controlling hyperbolic van der Waals polaritons," *Photon. Res.* **10**, B14-B22 (2022).

36. Y. Wu, J. Liu, W. Yu, T. Zhang, H. Mu, G. Si, Z. Cui, S. Lin, B. Zheng, C.-W. Qiu, H. Chen, and Q. Ou, "Monolithically Structured van der Waals Materials for Volume-Polariton Refraction and Focusing," *ACS Nano* **18** (26), 17065-17074 (2024).
37. See Supplementary Material at [link] for detailed derivations of the Minkowski-space formalism for hyperbolic media, the Descartes equation for hyperbolic lenses in Minkowski geometry, the construction of the lens welding region, the calculation of the numerical aperture and power balance, the derivation of the lens Green's and transfer functions, and which includes Ref. [38,41].
38. H. Goldstein, C. P. Poole, J. L. Safko, "Classical Mechanics," 3rd ed, (Pearson Inc., 2001).
39. M. P. d. Carmo, "Differential Geometry of Curves and Surfaces," 1st ed, (Prentice-Hall, Englewood Cliffs, NJ, 1976).
40. W. Ma, G. Hu, D. Hu, R. Chen, T. Sun, X. Zhang, Q. Dai, Y. Zeng, A. Alù, et al., "Ghost hyperbolic surface polaritons in bulk anisotropic crystals," *Nature* **596**, 362–366 (2021).
41. J. D. Jackson, Classical Electrodynamics, 3rd ed. (Wiley, New York, 1998).
42. A. S. Potemkin, A. N. Poddubny, P. A. Belov, and Y. S. Kivshar, "Green function for hyperbolic media," *Phys. Rev. A* **86**, 023848 (2012).
43. J. D. Caldwell, L. Lindsay, V. Giannini, I. Vurgaftman, T. L. Reinecke, S. A. Maier, O. J. and Glembocki, "Low-loss, infrared and terahertz nanophotonics using surface phonon polaritons," *Nanophotonics* **4** (1), 44-68 (2015).
44. S. Foteinopoulou, G. C. R. Devarapu, G. S. Subramania, S. Krishna, and D. Wasserman, "Phonon-polaritonics: enabling powerful capabilities for infrared photonics," *Nanophotonics* **8**, 12, 2129-2175 (2019).
45. S. Kim, S. G. Menabde, V. W. Brar, and M. S. Jang, "Functional Mid-Infrared Polaritonics in van der Waals Crystals," *Adv. Opt. Mat.* 8, 5 1901194 (2020).
46. G. Álvarez-Pérez, K. V. Voronin, V. S. Volkov, P. Alonso-González, and A. Y. Nikitin, "Analytical approximations for the dispersion of electromagnetic modes in slabs of biaxial crystals," Phys. Rev. B 100, 235408 (2019).
47. H. Teng, N. Chen, H. Hu, F. J. García de Abajo, Q. Dai, "Steering and cloaking of hyperbolic polaritons at deep-subwavelength scales," *Nat. Commun*. **15**, 4463 (2024).
48. P. Li, M. Lewin, A. V. Kretinin, J. D. Caldwell, K. S. Novoselov, T. Taniguchi, K. Watanabe, F. Gaussmann, and T. Taubner, "Hyperbolic phonon-polaritons in boron nitride for near-field optical imaging and focusing," *Nat. Commun.* **6**, 7507 (2015).

# Supplementary Material for “Minkowski-Space Modeling of Hyperbolic Lenses”

Enrico Maria Renzi[1,2], Simon Yves[1], Sveinung Erland[3], Diana Strickland[2], Eitan Bachmat[4], Andrea Alù[1,2]

*[1]Photonics Initiative, Advanced Science Research Center, City University of New York, New York, NY,10031, USA*
*[2]Physics Program, The Graduate Center, City University of New York, NY, 10026, USA*
*[3]Department of Mechanical Engineering and Maritime Studies, Western Norway University of Applied Sciences, 5528 Haugesund, Norway*
*[4]Department of Computer Science, Ben-Gurion University of the Negev, Beer Sheva, 84105, Israel*
aalu@gc.cuny.edu

## 1. Minkowski space formalism for hyperbolic media

### 1.1 *Effective Euclidean and Minkowski metrics in hyperbolic media*

Consider the propagation of a plane wave $\mathbf{E}_0 \exp(i\mathbf{k}\cdot\mathbf{r} - i\omega t)$ in a uniaxial medium characterized by relative permittivity tensor $\hat{\epsilon} = \mathrm{diag}\left(\epsilon_{xx}, \epsilon_{yy}, \epsilon_{zz}\right)$. Combining Ampere's and Faraday's law,

$$\nabla \times \mathbf{E} = ik_0 Z_0 \mathbf{H}\,, \tag{S.1}$$

$$\nabla \times \mathbf{H} = -i\frac{k_0}{Z_0}\hat{\epsilon}\mathbf{E}\,, \tag{S.2}$$

where $k_0 = \omega/c$ and $Z_0 = \sqrt{\mu_0/\epsilon_0}$ one obtains the homogeneous Helmholtz equation

$$\nabla \times \nabla \times \mathbf{E} + k_0^2 \hat{\epsilon}\mathbf{E} = 0\,. \tag{S.3}$$

where $k_0 = \omega / c$. Let $\mathbf{E}_0 = \left[E_x, E_y, E_z\right]^T$, $\mathbf{k} = \left[k_x, k_y, k_z\right]^T$, and $k^2 = k_x^2 + k_y^2 + k_z^2$. Under these definitions Eq. (S.3) can be rewritten in matrix form as:

$$\Lambda(\mathbf{k})\mathbf{E}_0 = 0 \tag{S.4}$$

where

$$\Lambda(\mathbf{k}) = \begin{pmatrix} k_x^2 - k^2 + k_0^2\epsilon_{xx} & k_x k_y & k_x k_z \\ k_x k_y & k_y^2 - k^2 + k_0^2\epsilon_{yy} & k_y k_z \\ k_x k_z & k_y k_z & k_z^2 - k^2 + k_0^2\epsilon_{zz} \end{pmatrix}. \tag{S.5}$$

Nontrivial solutions exist only when the determinant of the system vanishes, yielding the dispersion equation

$$F(k_x, k_y, k_z, k_0) = \det\left[\Lambda(\mathbf{k})\right] = 0\,, \tag{S.6}$$

where

$$\begin{aligned} F(k_x, k_y, k_z, k_0) = {} & \epsilon_{xx}\epsilon_{yy}\epsilon_{zz}k_0^4 - k_0^2\left[\epsilon_{xx}\epsilon_{yy}\left(k_x^2 + k_y^2\right) + \epsilon_{xx}\epsilon_{zz}\left(k_x^2 + k_z^2\right) + \epsilon_{yy}\epsilon_{zz}\left(k_y^2 + k_z^2\right)\right] \\ & + \epsilon_{xx}\left(k_x^4 + k_x^2k_y^2 + k_x^2k_z^2\right) + \epsilon_{yy}\left(k_y^4 + k_x^2k_y^2 + k_y^2k_z^2\right) + \epsilon_{zz}\left(k_z^4 + k_x^2k_z^2 + k_y^2k_z^2\right), \end{aligned} \tag{S.7}$$

defines the quartic Fresnel surface in k-space for a general diagonal anisotropic medium. This equation specifies the allowed wave vectors for an arbitrary field $\mathbf{E}_0$, with no polarization assumed at this stage. Eq. (S.7) does not admit closed-form solutions in general and cannot be reduced to a single quadratic form in $\mathbf{k}$.

To obtain analytical closed-form expressions, we restrict propagation to the principal planes, where polarization decouples and the quartic determinant factorizes. For a uniaxial, hyperbolic medium with $\hat{\epsilon} = \mathrm{diag}\left(-\epsilon_{\parallel}, \epsilon_{\perp}, \epsilon_{\perp}\right)$, $\epsilon_{\parallel}\epsilon_{\perp} > 0$, for which the optical axis is oriented along *x*. Restricting the propagation onto the *xy* plane, such that $\mathbf{k} = (k_x, k_y, 0)^T$, the dispersion relation separates into two quadratic surfaces:

$$k_x^2 + k_y^2 = \epsilon_{\perp} k_0^2 \tag{S.8}$$

and

$$\frac{k_x^2}{\epsilon_{\perp}} - \frac{k_y^2}{\epsilon_{\parallel}} = k_0^2 \tag{S.9}$$

for the ordinary waves and extraordinary waves, respectively. Therefore, because of Eq. (S.8) the ordinary wave experiences an isotropic medium, while Eq. (S.9) suggests that the extraordinary wave experiences anisotropic propagation. Both dispersion relations can be written compactly using an effective metric in reciprocal space. For the ordinary space, the dispersion rewrites as

$$\mathbf{k}^{\mathbf{T}} \cdot \hat{G}^{(o)} \cdot \mathbf{k} = k_0^2, \qquad \hat{G}^{(o)} = (\epsilon_\perp)^{-1} \operatorname{diag}(1,1), \tag{S.10}$$

which corresponds to an Euclidean metric in k-space. For the extraordinary wave, instead we have

$$\mathbf{k}^{\mathbf{T}} \cdot \hat{G}^{(e)} \cdot \mathbf{k} = k_0^2, \qquad \hat{G}^{(e)} = \operatorname{diag}\left(-1/\epsilon_\perp, 1/\epsilon_\parallel\right), \tag{S.11}$$

which has the structure of a Minkowski metric. Since wave propagation relates wavevector and position through the phase $\mathbf{k} \cdot \mathbf{r}$, the natural real-space metric is the inverse of the k-space metric. This reciprocal relation ensures that the dot products in real space reproduce the same geometric weights that appear in momentum space. Then the infinitesimal distance in real space is defined as:

$$ds^2 = d\mathbf{r}^T \cdot \hat{M} \cdot d\mathbf{r}, \qquad \hat{M} = \left(\hat{G}\right)^{-1}. \tag{S.12}$$

Explicitly, for the ordinary wave the real space metric is

$$ds_{(o)}^2 = \epsilon_\perp \left(dx^2 + dy^2\right), \tag{S.13}$$

while for the extraordinary wave in the same plane, the metric becomes

$$ds_{(e)}^2 = \epsilon_\perp dx^2 - \epsilon_\parallel dy^2 . \tag{S.14}$$

This metric determines how dot product and distances are computed and therefore controls the geometry of rays, which follow geodesics of the corresponding metric.

An important observation is that ordinary and extraordinary waves occupy orthogonal polarization subspaces for waves propagating in any plane parallel to the principal optical axis as we demonstrate below.

For example, in the *x-y* plane the extraordinary wave is a TM-polarized wave with magnetic field along *z*,

$$\mathbf{H}_e = (0,0,H_z), \qquad \mathbf{E}_e = (E_x, E_y, 0), \tag{S.15}$$

while the ordinary wave is TE-polarized with fields

$$\mathbf{H}_o = (H_x, H_y, 0), \qquad \mathbf{E}_o = (0,0,E_z). \tag{S.16}$$

Therefore, ordinary and extraordinary waves live in planes of polarizations governed by their own metric: the Euclidean metric for the ordinary mode and the Minkowski-type metric for the extraordinary mode. Because the modes do not mix, the two geometries coexist without interference.

This discussion can be extended to biaxial media, for which $\hat{\epsilon} = \operatorname{diag}\left(\epsilon_{xx}, \epsilon_{yy}, \epsilon_{zz}\right)$ with all components distinct. The full 3D dispersion, i.e., Eq. (S.7), no longer factorizes into two quadratics, and in general cannot be written using a single quadratic metric. However, when propagation is restricted along the principal planes ( $k_x = 0$ , $k_y = 0$ , or $k_z = 0$ ), the quartic again factorizes into two branches. For example, in the *x-y* plane one finds

$$\left(\frac{k_x^2 + k_y^2}{\epsilon_{zz}} - k_0^2\right)\left(\frac{k_x^2}{\epsilon_{yy}} + \frac{k_y^2}{\epsilon_{xx}} - k_0^2\right) = 0 , \tag{S.17}$$

Which reproduce a Euclidean ordinary metric and an anisotropic extraordinary metric exactly as in the uniaxial case. The same occurs in the *x-z* and *y-z* planes. Only in these planes the two polarizations remain unmixed, allowing two independent metric descriptions.

Away from principal planes, biaxial media mix polarizations and the dispersion surface remains quartic. Because the wave does not remain confined to orthogonal polarization subspaces, it is impossible to define a single quadratic metric governing propagation. In other words, the effective geometry is no longer flat: a quadratic metric corresponds to a flat geometric structure [20], while the full quartic Fresnel surface of a biaxial medium cannot be mapped to such a space. Hence, the metric description is exact only in uniaxial media along the $\parallel$ axis, or in biaxial media restricted to propagation along principal planes where the field polarizations remain decoupled.

These considerations apply broadly to *any* hyperbolic platform in which the field naturally decomposes into an isotropic branch and an anisotropic (hyperbolic) branch. Whenever the polarizations are separable, each branch is governed by its own effective metric, offering a universal geometric framework to describe wave propagation in hyperbolic media.

### *1.2 Phase and ray direction in hyperbolic media*

We derive the phase and ray propagation directions for plane waves in an unbounded hyperbolic medium and demonstrate that, while misaligned in Euclidean space (Cartesian coordinates), these quantities become parallel in Minkowski geometry (hyperbolic coordinates).

Consider an unbound, non-dispersive and lossless uniaxial hyperbolic medium with permittivity tensor

$$\hat{\epsilon} = \mathrm{diag}(-\epsilon_{xx}, \epsilon_{yy}, \epsilon_{yy}), \tag{S.18}$$

where $\epsilon_{xx}\epsilon_{yy} > 0$. Then the propagation of extraordinary (transverse magnetic) waves with wave vector $\mathbf{k} = \left[k_x, k_y, 0\right]^T$ is governed by the normalized dispersion equation

$$\frac{k_x^2}{\epsilon_{yy}} - \frac{k_y^2}{\epsilon_{xx}} = 1 \tag{S.19}$$

Then, the in-plane wave vector is

$$\boldsymbol{k} = \left[\sqrt{\epsilon_{yy} + \frac{\epsilon_{yy}}{\epsilon_{xx}} k_y^2}, k_y\right]^T, \tag{S.20}$$

and the phase propagation angle with respect to the $x$ axis is [Fig. S1a]:

$$\alpha = \tan^{-1}\left(\frac{k_y}{k_x}\right) = \tan^{-1}\left[k_y\left(\epsilon_{yy} + \epsilon_{yy}\epsilon_{xx}{}^{-1}k_y^2\right)^{-\frac{1}{2}}\right]. \tag{S.21}$$

We define a direction-dependent effective index $n(\alpha)$ such that the wavevector components satisfy $k_x = n(\alpha)\cos\alpha$ and $k_y = n(\alpha)\sin\alpha$. This is equivalent to defining a refractive index for isotropic media, with the difference that in this case anisotropy is encoded inside the scalar $n(\alpha)$. The expression for $n(\alpha)$ is obtained inserting $k_x$ and $k_y$ into Eq. (S.19), and is given by:

$$n(\alpha) = \sqrt{\epsilon_{xx}\epsilon_{yy}}\left(\epsilon_{xx}\cos^2\alpha - \epsilon_{yy}\sin^2\alpha\right)^{-1/2} \tag{S.22}$$

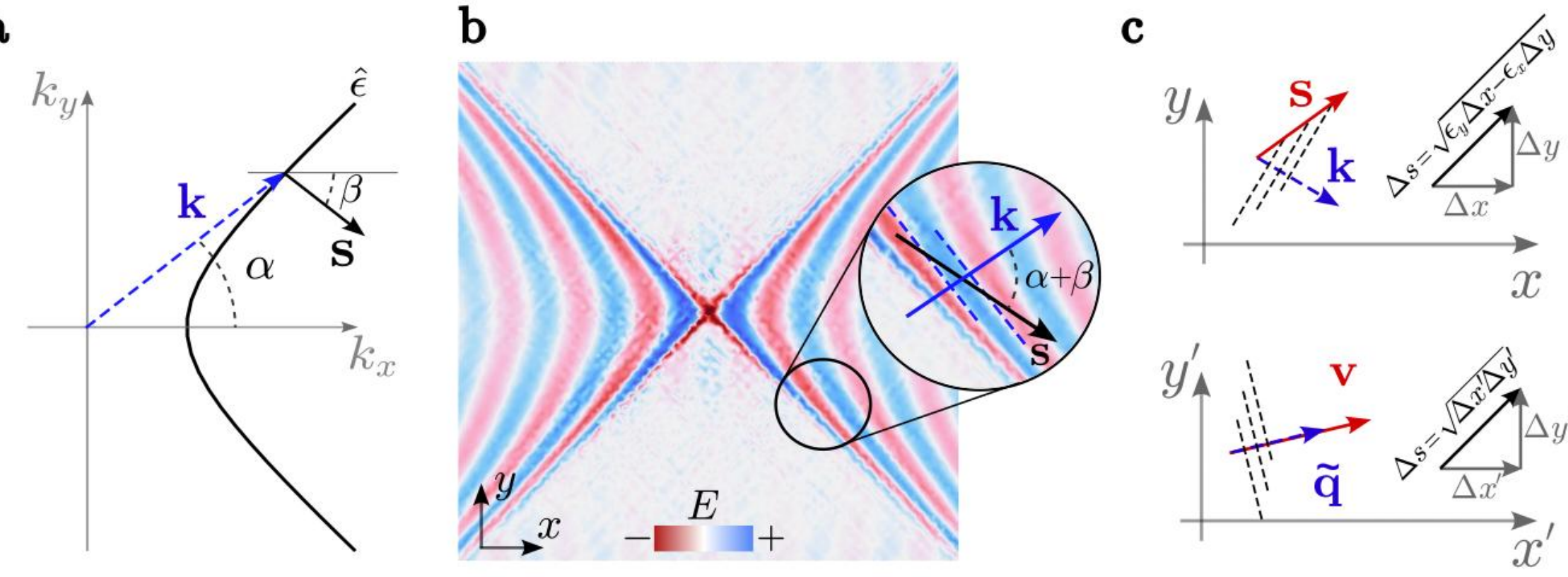

**Figure S1: Wave propagation and coordinate transformation in hyperbolic media.** (a) Wave propagation in hyperbolic media is governed by hyperbolic iso-frequency contours. The phase propagation direction ($\alpha$) differs from the power flow direction ($\beta$). (b) Full-wave simulation of a point dipole emitting in a non-dispersive hyperbolic medium with permittivity $\hat{\epsilon} = \mathrm{diag}([-1,1,1])$, highlighting the angular mismatch between wavefronts and energy flow directions (inset). (c) To simplify this misalignment, a transformation into Minkowski coordinates is applied. In this new frame, the anisotropy is embedded into the metric. Then the phase is calculated as $\phi = \tilde{\mathbf{q}} \cdot \hat{M}' \mathbf{v}$, where $\tilde{\mathbf{q}}$ is the group velocity vector and $\mathbf{v}$ the ray oriented path.

The asymptotic angle $\tan^{-1}\sqrt{m} = \tan^{-1}\sqrt{\epsilon_{xx}/\epsilon_{yy}}$ defines the aperture of the hyperbolic iso-frequency contours. As $\alpha$ approaches this angle, the effective index $n(\alpha)$ diverges, indicating increasingly confined wave propagation along the asymptotic direction.

Due to the medium's anisotropy, the power flow direction angle $\beta = \tan^{-1}\left(S_y / S_x\right)$ differs from the phase velocity direction $\alpha$ for a given plane wave [Fig. S1a-b]. The angle $\beta$ is found calculating the normal to the hyperbolic contours (S.19), which is given by the unit vector

$$\hat{\mathbf{s}} = \left(1 + \frac{k_y^2 \epsilon_{yy}}{\epsilon_{xx}^2 - k_y^2 \epsilon_{xx}}\right)^{-\frac{1}{2}} \left[1, -k_y \frac{\epsilon_{yy}}{\epsilon_{xx}} \left(\epsilon_{yy} + \frac{\epsilon_{yy}}{\epsilon_{xx}} k_y^2\right)^{-\frac{1}{2}}\right]^T. \tag{S.23}$$

Then, defining the unit vector components as $\hat{s}_x = \cos\beta$ and $\hat{s}_y = \sin\beta$, from the equation above we obtain the wave vector in Eq. (S.20) as a function of the power flow angle $\beta$:

$$\mathbf{k} = \left[k_x(\beta), k_y(\beta)\right]^T = \frac{1}{\sqrt{\epsilon_{yy} \cos^2\beta - \epsilon_{xx} \sin^2\beta}} \left[\epsilon_{yy} \cos\beta, -\epsilon_{xx} \sin\beta\right]^T, \tag{S.24}$$

where $-\pi/2 < \beta < \pi/2$. Notice that $\beta$ differs from $\alpha$. However, comparing the wave vector expression in equation (S.24) and (S.20), we obtain:

$$\tan\beta = -\frac{\epsilon_{yy}}{\epsilon_{xx}} \tan\alpha \tag{S.25}$$

Therefore, power flow and phase are in general non-parallel, except when $k_y = 0$.

### *1.3 Phase accumulation in Euclidean geometry*

To compute the accumulated phase along a propagation direction $\mathbf{s}$, we evaluate the scalar product

$$\phi = \frac{2\pi}{\lambda} \mathbf{k} \cdot \mathbf{s} \tag{S.26}$$

where $\lambda = 2\pi c_0 / \omega$ and $\omega$ is the angular frequency. In conventional Cartesian coordinates, this product is simply defined by the identity metric, meaning it can be written as

$$\phi = \frac{2\pi}{\lambda} \mathbf{k}^T \cdot \begin{pmatrix} 1 & 0 \\ 0 & 1 \end{pmatrix} \mathbf{s} \tag{S.27}$$

Now, considering a straight ray $\mathbf{s} = [\Delta x, \Delta y]^T$, where $\Delta x = x - x_0$ and $\Delta y = y - y_0$ ($x_0$ and $y_0$ are the origin point of the ray), such that the propagation angle is $\beta = \tan^{-1}\left(\Delta y / \Delta x\right)$, we substitute the expression for $\mathbf{k}(\beta)$ in (S.24), obtaining:

$$\mathbf{k} = (k_x, k_y) = \frac{1}{\sqrt{\epsilon_{yy} \Delta x^2 - \epsilon_{xx} \Delta y^2}} \left[\epsilon_{yy} \Delta x, -\epsilon_{xx} \Delta y\right]^T, \tag{S.28}$$

It follows directly that $\mathbf{k}$ and $\mathbf{s}$ are not parallel, as their cross product is nonzero:

$$\mathbf{k} \times \mathbf{s} = \hat{\mathbf{z}} \frac{(\epsilon_{xx} + \epsilon_{yy}) \Delta x \Delta y}{\sqrt{\epsilon_{yy} \Delta x^2 - \epsilon_{xx} \Delta y^2}}, \tag{S.29}$$

and, using (S.26), the accumulated phase is

$$\phi = \frac{2\pi}{\lambda} \frac{\epsilon_{yy} \Delta x^2 - \epsilon_{xx} \Delta y^2}{\sqrt{\epsilon_{yy} \Delta x^2 - \epsilon_{xx} \Delta y^2}} \tag{S.30}$$

Which can be simplified as

$$\phi = \frac{2\pi}{\lambda} \sqrt{\epsilon_{yy} \Delta x^2 - \epsilon_{xx} \Delta y^2} \tag{S.31}$$

Notably, this expression resembles the generalized Minkowski distance in a (1+1)-dimensional spacetime [16], which describes the path traveled by a massless particle governed by the Klein-Gordon equation.

### *1.4 Phase accumulation in Minkowski geometry*

Indeed, for transverse magnetic waves propagating in the $(x, y)$ plane with magnetic field $\mathbf{H} = \hat{\mathbf{z}}H_z$, it can be shown that the wave equation takes an analogous form [21]:

$$\frac{1}{\epsilon_{yy}}\frac{\partial^2}{\partial x^2}H_z - \frac{1}{\epsilon_{xx}}\frac{\partial^2}{\partial y^2}H_z = -\frac{\lambda^2}{(2\pi)^2}H_z \tag{S.32}$$

This analogy becomes explicit by rescaling the spatial coordinates according to the permittivity components $\epsilon_{xx}$ and $\epsilon_{yy}$, defining a stretched Minkowski space that incorporates the anisotropic material properties. Within this framework, we introduce a metric tensor $\hat{M} = \mathrm{diag}([\epsilon_{yy}, -\epsilon_{xx}])$. This allows us to distinguish between the covariant momentum vector $\mathbf{q}$ and its contravariant counterpart $\tilde{\mathbf{q}}$. The covariant vector $\mathbf{q}$ is the momentum vector, whereas the contravariant vector $\tilde{\mathbf{q}}$ determines the direction of energy flow. Unlike in Euclidean space, vectors and covectors are not equivalent in a hyperbolic medium; instead, $\mathbf{q}$ must first be converted to its contravariant form $\tilde{\mathbf{q}}$ through the metric, having,

$$\phi = \mathbf{q}\cdot\mathbf{v} = \tilde{\mathbf{q}}\cdot\hat{M}^{-1}\mathbf{s} \tag{S.33}$$

where $\tilde{\mathbf{q}}$ is parallel to the ray displacement $\mathbf{s}$ (as in isotropic media).

### *1.5 Light-cone coordinate transformation*

We begin from the definition of phase in Euclidean expressed as in Eq. (S.31) is

$$\phi(x, y) = k_0\sqrt{\epsilon_{yy}}\sqrt{x^2 - my^2} \tag{S.34}$$

It is now convenient to perform a change of coordinates, to further simplify the calculations. Given the hyperbolic contour aperture $\sqrt{m} = \sqrt{\epsilon_{xx}/\epsilon_{yy}}$, we stretched and rotate the system as follows:

$$\begin{aligned} x &= \tfrac{1}{\sqrt{2}}(x' + y'), \\ y &= \tfrac{1}{\sqrt{2m}}(y' - x'). \end{aligned} \tag{S.35}$$

We note that this transformation generalizes the transformation to light-cone (Dirac) coordinates [30]. This change of variables maps the physical problem from the $(x, y)$ plane, where geometry is tied to the permittivity-tensor parameters, onto a plane $(x' > 0, y' > 0)$ [Fig. S1c], in which the calculations become independent of the specific aperture $\sqrt{m}$. Under this transformation, the phase in Eq. (S.34) becomes

$$\phi(x', y') = k_0\sqrt{\epsilon_{yy}}\sqrt{2x'y'} \tag{S.36}$$

Phase accumulation is a scalar, and hence invariant under any change of coordinates. Therefore, the momentum vector $\mathbf{q}$ is simply the gradient of phase $\nabla_{\mathbf{r}'}\phi$, giving

$$q_{x'} = \frac{\partial\phi}{\partial x'} = k_0\sqrt{\frac{\epsilon_{yy}}{2}}\frac{y'}{\sqrt{x'y'}}, \tag{S.37}$$

and

$$q_{y'} = \frac{\partial\phi}{\partial y'} = k_0\sqrt{\frac{\epsilon_{yy}}{2}}\frac{x'}{\sqrt{x'y'}} \tag{S.38}$$

Under this construction, the anisotropy is lifted and we obtain a direction- and k-space aperture independent effective index $n_e = \sqrt{\epsilon_{yy}/2}$. In this geometry the metric is

$$\hat{M}' = \left(\hat{M}'\right)^{-1} = \begin{pmatrix} 0 & 1 \\ 1 & 0 \end{pmatrix}. \tag{S.39}$$

so that the generalized distance between two points is

$$\left(\Delta s\right)^2 = \begin{pmatrix} \Delta x \\ \Delta y \end{pmatrix}^T \begin{pmatrix} 0 & 1 \\ 1 & 0 \end{pmatrix} \begin{pmatrix} \Delta x \\ \Delta y \end{pmatrix} = 2\Delta x' \Delta y' \tag{S.40}$$

where $\Delta x' = x' - x_0'$ and $\Delta y' = y' - y_0'$, are component of the displacement vectors $\mathbf{v} = \left(\Delta x', \Delta y'\right)^T$, and that the phase accumulation is

$$\phi = \mathbf{q} \cdot \mathbf{v} = k_0 \sqrt{2\epsilon_{yy} \Delta x' \Delta y'} \tag{S.41}$$

where

$$\mathbf{q} = k_0 n_e \frac{1}{\sqrt{\Delta x' \Delta y'}} \begin{pmatrix} \Delta y' \\ \Delta x' \end{pmatrix}, \qquad \tilde{\mathbf{q}} = k_0 n_e \frac{1}{\sqrt{\Delta x' \Delta y'}} \begin{pmatrix} \Delta x' \\ \Delta y' \end{pmatrix} \tag{S.42}$$

Then in this new frame, Eq. (S.41) is a ready to use equation to calculate the phase along any straight path with horizontal displacement $\Delta x'$ and vertical displacement $\Delta y'$. It is instructive to observe that the phase accumulation equals $\phi = k_0 \sqrt{\epsilon_{yy}} \sqrt{\mathbf{v} M' \mathbf{v}} = k_0 \sqrt{\epsilon_{yy}} \mathbf{v}_{M'}$, where $\mathbf{v}_{M'}$ is the length of the displacement vector $\mathbf{v}$ under the metric $M'$.

Notice that even in this frame of reference, the wave vector $\mathbf{q}$ and ray vector $\tilde{\mathbf{q}}$ are symmetric about the diagonal by construction, but $\tilde{\mathbf{q}}$ is simply a rescaled version of the ray direction, equivalently to the Euclidean case. We conclude pointing out that the wave vector magnitude is given by:

$$\left|\mathbf{q}\right|^2 = \left|\tilde{\mathbf{q}}\right|^2 = \mathbf{q}^T \cdot \hat{M}' \mathbf{q} = 2 q_{x'} q_{y'} = k_0 \epsilon_{yy} . \tag{S.43}$$

### *1.6 Maxwell equations for TM waves in Minkowski space*

We restrict the analysis to TM waves, for which $\mathbf{H} = H_z \hat{\mathbf{z}}$. Starting from Eq. (S.1), the electric fields components in Euclidean space can be expressed as:

$$E_x = \frac{iZ_0}{k_0 \epsilon_{xx}} \frac{\partial H_z}{\partial y}, \qquad E_y = -\frac{iZ_0}{k_0 \epsilon_{yy}} \frac{\partial H_z}{\partial x} . \tag{S.44}$$

Switching to generalized light-cone coordinates we have:

$$\frac{\partial}{\partial x} = \frac{1}{\sqrt{2}} \left( \frac{\partial}{\partial x'} + \frac{\partial}{\partial y'} \right), \qquad \frac{\partial}{\partial y} = \sqrt{\frac{m}{2}} \left( \frac{\partial}{\partial y'} - \frac{\partial}{\partial x'} \right), \tag{S.45}$$

from which we obtain:

$$E_x = \frac{iZ_0 \sqrt{m}}{k_0 \epsilon_{xx} \sqrt{2}} \left( \frac{\partial H_z}{\partial y'} - \frac{\partial H_z}{\partial x'} \right), \qquad E_y = \frac{-iZ_0}{k_0 \epsilon_{yy} \sqrt{2}} \left( \frac{\partial H_z}{\partial x'} + \frac{\partial H_z}{\partial y'} \right). \tag{S.46}$$

Then we connect the electric field in Euclidean coordinates to those in Minkowski space:

$$\begin{aligned} E_{x'} &= \frac{\partial x}{\partial x'} E_x + \frac{\partial y}{\partial x'} E_y = \frac{1}{\sqrt{2}} E_x - \frac{1}{\sqrt{2m}} E_y , \\ E_{y'} &= \frac{\partial x}{\partial y'} E_x + \frac{\partial y}{\partial y'} E_y = \frac{1}{\sqrt{2}} E_x + \frac{1}{\sqrt{2m}} E_y \end{aligned} \tag{S.47}$$

Inserting the expressions in Eq. (S.46) we obtain:

$$E_{x'} = \frac{iZ_0}{k_0} \frac{1}{\sqrt{\epsilon_{xx} \epsilon_{yy}}} \frac{\partial H_z}{\partial y'}, \qquad E_{y'} = -\frac{iZ_0}{k_0} \frac{1}{\sqrt{\epsilon_{xx} \epsilon_{yy}}} \frac{\partial H_z}{\partial x'} . \tag{S.48}$$

Comparing Eqs. (S.44) with Eqs. (S.48), we observe that the transformation to light-cone coordinates maps wave propagation in the original hyperbolic medium to an equivalent problem in an isotropic medium with effective wave impedance $Z_{eff} = Z_0 / \sqrt{\epsilon_{xx}\epsilon_{yy}}$ . We look for plane wave solutions, then $H_z = H_0 \exp(i\mathbf{q}' \cdot \mathbf{r}')$ , then:

$$\begin{aligned} \frac{\partial H_z}{\partial x'} &= i \frac{q_{y'}}{2} H_z \\ \frac{\partial H_z}{\partial y'} &= i \frac{q_{x'}}{2} H_z \end{aligned} \tag{S.49}$$

which inserted in Eq. (S.49) give:

$$E_{x'} = -\frac{Z_{eff}}{k_0}\frac{q_{x'}}{2} H_z, \qquad E_{y'} = \frac{Z_{eff}}{k_0}\frac{q_{y'}}{2} H_z . \tag{S.50}$$

## 2. Descartes equation for hyperbolic lenses in Minkowski geometry

### *2.1 Collimating lens construction*

In this section, we present the construction of a lensing interface between two hyperbolic media characterized by permittivity tensors $\hat{\epsilon}_1 = \mathrm{diag}(-\epsilon_{1,xx}, \epsilon_{1,yy})$ and $\hat{\epsilon}_2 = \mathrm{diag}(-\epsilon_{2,xx}, \epsilon_{2,yy})$ using the Minkowski geometry tools presented previously. For this analysis, we assume the lens optical axis is parallel to the $x$-axis and that the two media are homogenous and unbound in the $z$ direction [Fig. S2a, rightmost inset].

In Cartesian coordinates, the shape of a lens that focuses radiation from a point source placed at the origin (0,0) to an image plane at infinity, with the lens positioned at distance $f$ and its optical axis parallel to the $x$-axis, has been derived previously in [33] and is given by the hyperbolic equation

$$\left(y - \frac{\sqrt{\epsilon_{2,xx}}}{\sqrt{\epsilon_{1,xx}}+\sqrt{\epsilon_{2,xx}}} f\right)^2 + \frac{\epsilon_{1,yy}}{\epsilon_{1,xx}-\epsilon_{2,xx}} x^2 = \frac{\epsilon_{1,xx}}{\left(\sqrt{\epsilon_{1,xx}}+\sqrt{\epsilon_{2,xx}}\right)^2} f^2, \tag{S.51}$$

however, in our work, we take advantage of Minkowski space, and solve the problem in the transformed space assuming that the lens is constructed between materials with indexes $n_1 = \sqrt{\epsilon_{1,yy}/2}$ and $n_2 = \sqrt{\epsilon_{2,yy}/2}$ [Fig. S2a] simplifying the derivation. Without loss of generality, we assume hyperbolic media with the same apertures, such that

$$m = \frac{\epsilon_{1,xx}}{\epsilon_{1,yy}} = \frac{\epsilon_{2,xx}}{\epsilon_{2,yy}}, \tag{S.52}$$

so that the phase in both media is given by (S.41). In this frame, the optical axis of the lens aligns with the $y' = x'$ line, intersecting the lens at the coordinate $f' = (x_f', y_f')$. Then using conventional Descartes techniques, but in Minkowski geometry, we equate the phase along the optical axis

$$\phi_{OA} = \frac{1}{2}\sqrt{2\epsilon_{y,1}} \tag{S.53}$$

with the phase accumulated along any other path that first intersects the lens profile and then hits the image plane (at the antidiagonal) parallel to the optical axis [Fig. S2a], for which the phase is:

$$\phi_{OB} + \phi_{BC} = \sqrt{2\epsilon_{y,1}}\sqrt{x'y'} + \sqrt{2\epsilon_{y,2}}\frac{1-x'-y'}{2}. \tag{S.54}$$

We obtain the lens characteristic equation

$$F(x', y') = \frac{1}{2}\sqrt{2\epsilon_{y,1}} - \sqrt{2\epsilon_{y,1}}\sqrt{x'y'} - \sqrt{2\epsilon_{y,2}}\frac{1-x'-y'}{2} = 0 \tag{S.55}$$

Defining the index contrast

$$T = \sqrt{\frac{\epsilon_{2,yy}}{\epsilon_{1,yy}}}, \tag{S.56}$$

Eq. (S.54) simplifies to

$$F(x', y') = 1 - T - 2\sqrt{x'y'} + T(x' + y') = 0. \tag{S.57}$$

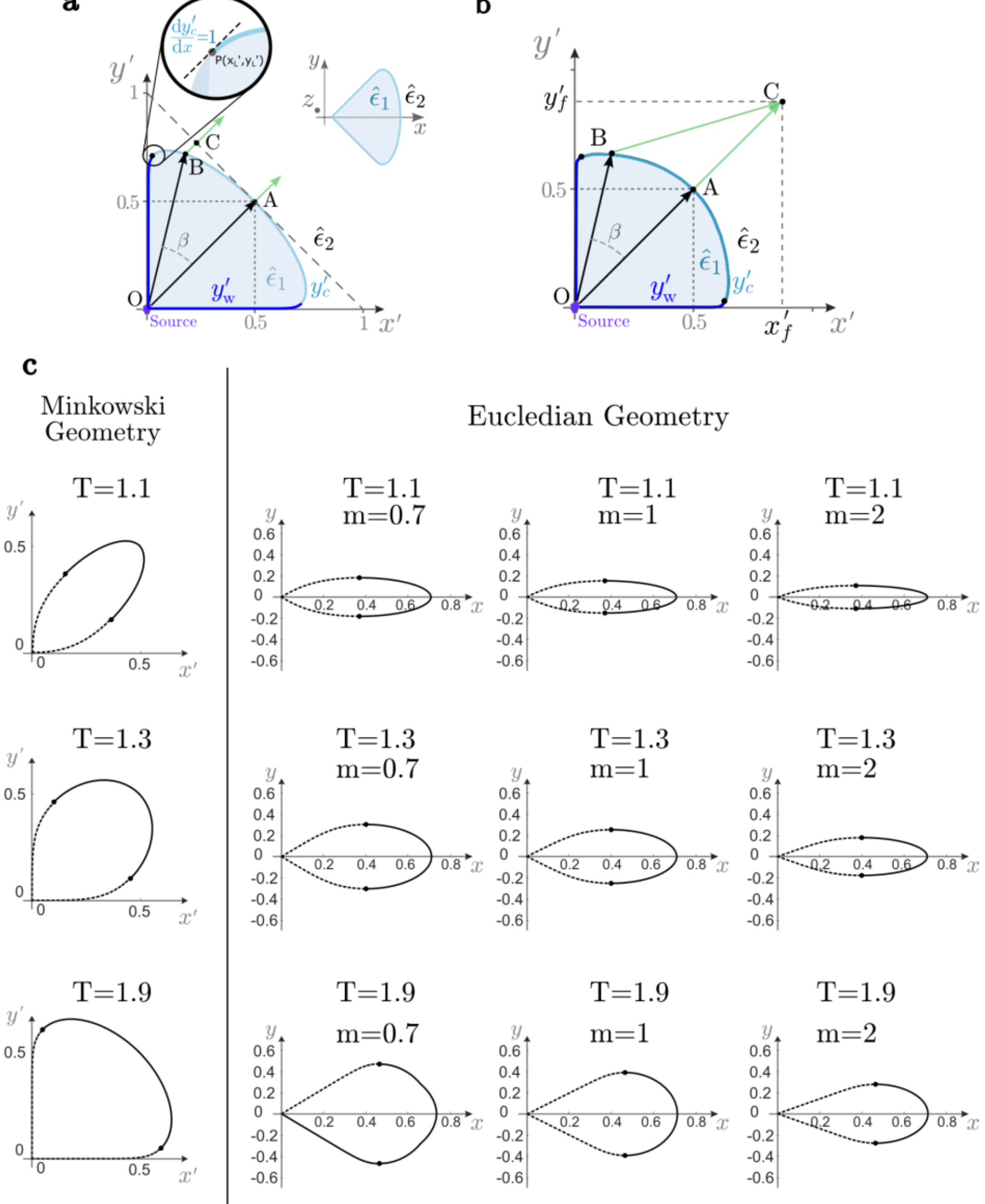

**Figure S2: Hyperbolic lens design via Minkowski geometry.** (a) Construction of the hyperbolic lens by transforming from Cartesian coordinates (inset) to Minkowski coordinates. The lens comprises two segments: a collimating part ($y_c'$, solid blue) refracting rays toward infinity, and a welding region ($y_w'$, dark blue) smoothly redirecting marginal rays at the lens edge, reducing diffraction. The lens shape is obtained by applying Descartes' method in Minkowski geometry. (b) Extension of the lens construction to a converging lens with finite focal point $(x_f', y_f')$. (c) Various configurations of material contrast $T=\sqrt{\epsilon_{2,yy}/\epsilon_{1,yy}}$ and k-space aperture $m_1=m_2=m=\epsilon_{1,xx}/\epsilon_{1,yy}$, for (left) light-cone coordinates, independent of k-space aperture, and (right) in Euclidean coordinates. The collimating section (black solid line) and anti-reflection section (black dashed line) are separated by the line-of-sight point (black solid dot).

Solving the equation above, we find the expression for the lens:

$$y_c'(x') = \left(\frac{\sqrt{x'}+\sqrt{(T-1)[T-x'(T+1)]}}{T}\right)^2 \tag{S.58}$$

which is valid for $x_L' < x' < 1/2$ (light blue solid line in Fig. S2a). The coordinate $P_L' = (x_L', y_L')$ defines the line of sight of the lens [Fig. S2a, circular inset], which is the point beyond which captured rays are no more collimated. Because the lens interface is parallel to the line $y = x$ at the line-of-sight point, this coordinate is found solving:

$$\left(\frac{dy_c{}'}{dx'}\right)_{x'=x_L{}'} = 1, \tag{S.59}$$

which yields

$$x_L{}' = \frac{1}{2\left(T+T^2+(T+1)\sqrt{(T-1)(T+1)}\right)} \tag{S.60}$$

Then, transforming back to Cartesian coordinates, we obtain the lens profile:

$$y_c = \sqrt{\frac{(T-1)\left(1-T+2\sqrt{2}xT-2(T+1)x^2\right)}{2m}}, \quad x_L < x < 1/\sqrt{2} \tag{S.61}$$

where the line-of-sight coordinates are

$$P_L = (x_L, y_L) = \left(\frac{T}{\sqrt{2}(1+T)}, \sqrt{\frac{T-1}{2m(T+1)}}\right) \tag{S.62}$$

Notice that the focal length in (S.61) is fixed to $f = 1/\sqrt{2}$, and can be rescaled to $\alpha f$, $\alpha > 0$, as needed for any particular design by rescaling the coordinates of the lens by the same factor.

### 2.2 Action for waves in hyperbolic media

In contrast to isotropic lenses, hyperbolic lenses exhibit an inverted curvature [Fig. S2]. This occurs because phase accumulates differently in hyperbolic media: whereas isotropic lenses compensate for the longer optical path of oblique rays, hyperbolic materials accumulate more phase along the optical axis and progressively less for oblique, large-momentum rays. As a result, the lens must flip its curvature so that both axial and marginal rays acquire the same total phase and converge constructively at the focus. This inversion follows from Fermat's principle adapted to hyperbolic geometry, where geodesics maximize the proper-time–like optical action, rather than minimize it. Before defining an equivalent action principle for hyperbolic media, we proceed deriving the Lagrangian, for TM waves propagating inside hyperbolic media, and from that the associated Action.

In the geometric optics limit, a wave packet is described by its position $\mathbf{r}$ and wave vector $\mathbf{k}$. In this framework, the action $S$, corresponds to the accumulated phase of the wave between the points $A$ and $B$ [1]:

$$S = \int_A^B d\phi = \int_A^B \mathbf{k}\cdot d\mathbf{r} \tag{S.63}$$

To define the Lagrangian $\mathcal{L}$, we must express the integral in terms of a single parameter. Following the definition (S.31) and the 1+1 space time mapping, we chose $x$ as the equivalent time coordinate and $y$ as the equivalent spatial coordinate. Then the differential displacement is

$$d\mathbf{r} = (dx, dy) = (1, \dot{y})dx \tag{S.64}$$

where $\dot{y} = dy/dx$. Hence:

$$S = \int_{x_A}^{x_B} (k_x + k_y\dot{y})dx \tag{S.65}$$

Following the definition of the Action, that is

$$S = \int_A^B \mathcal{L}(y, \dot{y}, x)dx\ , \tag{S.66}$$

we identify the Lagrangian for a TM mode propagating into a hyperbolic medium as

$$\mathcal{L}(y, \dot{y}, x) = k_x + k_y\dot{y}\ . \tag{S.67}$$

The ray follows the direction of the group velocity $\mathbf{s} = (s_x, s_y)$, for which $s_x = dx/dt$ and $s_y = dy/dt$. Then, we rewrite

$$\dot{y} = \frac{dy}{dx} = \frac{s_y dt}{s_x dt} = \frac{s_y}{s_x} \tag{S.68}$$

Equivalently, the group velocity is defined as the gradient of the frequency with respect to the wave vector:

$$\mathbf{s} = \nabla_{\mathbf{k}} \omega = \left( \frac{\partial \omega}{\partial k_x}, \frac{\partial \omega}{\partial k_y} \right) \tag{S.69}$$

Inside the hyperbolic medium, from Eq. (S.19), we get:

$$\omega = c\sqrt{\frac{k_x{}^2}{\epsilon_{yy}} - \frac{k_y{}^2}{\epsilon_{xx}}} . \tag{S.70}$$

Hence, using (S.70) in Eq. (S.69):

$$\dot{y} = \frac{\partial \omega / \partial k_y}{\partial \omega / \partial k_x} = \frac{-2k_y / \epsilon_{xx}}{2k_x / \epsilon_{yy}} . \tag{S.71}$$

Inverting the relation above, we find:

$$k_y = -k_x \frac{\epsilon_{xx}}{\epsilon_{yy}} \dot{y} . \tag{S.72}$$

Inserting into Eq. (S.70), and solving from $k_x$, we find:

$$k_x = \frac{\omega}{c} \sqrt{\epsilon_{yy}} \left( 1 - \frac{\epsilon_{xx}}{\epsilon_{yy}} \dot{y}^2 \right)^{-1/2} . \tag{S.73}$$

Inserting Eq. (S.72) and Eq. (S.73) into the Lagrangian definition in Eq. (S.67), we obtain:

$$\mathcal{L}(\dot{y}) = \frac{2\pi}{\lambda} \sqrt{\epsilon_{yy}} \sqrt{1 - \frac{\epsilon_{xx}}{\epsilon_{yy}} \dot{y}^2} . \tag{S.74}$$

We note that this result is fully consistent with what one would anticipate from the Minkowski-geometry formulation. Indeed, starting from the definition of the Minkowski distance in Eq. (S.31), we could have directly written the action as

$$S = \frac{2\pi}{\lambda} \int_A^B ds = \frac{2\pi}{\lambda} \int_A^B \sqrt{\epsilon_{yy} dx^2 - \epsilon_{xx} dy^2} = \frac{2\pi}{\lambda} \int_A^B \left( \epsilon_{yy} - \epsilon_{xx} \left( \frac{dy}{dx} \right)^2 \right)^{1/2} dx , \tag{S.75}$$

corroborating the advantage of working inside the framework of hyperbolic geometry.

### *2.3 Action maximization in hyperbolic media*

In the following, we prove rigorously that for this class of hyperbolic lenses the curvature must indeed be opposite to that of an isotropic lens, and that this result emerges directly from the Fermat action principle. Crucially, unlike in isotropic media, where the optical path is minimized, hyperbolic media require maximization of the action along the ray trajectory. This reversal fundamentally alters the geometric constraints imposed on the lens surface and leads naturally to the opposite curvature required for constructive focusing.

Stationary points of the action correspond to the system's geodesics, for which Fermat's principle requires that

$$\frac{d}{dx} \frac{\partial \mathcal{L}}{\partial \dot{y}} - \frac{\partial \mathcal{L}}{\partial y} = 0 . \tag{S.76}$$

Let the Lagrangian in Eq. (S.74) be written in the simplified notation:

$$\mathcal{L}(\dot{y}) = A\sqrt{1 - \alpha \dot{y}^2} , \tag{S.77}$$

where $A = 2\pi\sqrt{\epsilon_{yy}} / \lambda$ and $\alpha = \epsilon_{xx} / \epsilon_{yy}$ .

Stationary points of the action arise from the variational problem of solving $\delta S = 0$. Establishing whether they are maxima or minima amounts to examining the sign of the second variation $\delta^2 S \propto \partial^2 \mathcal{L} / \partial \dot{y}^2$ [38]. From Eq. (S.77), we have:

$$\frac{\partial^2 \mathcal{L}}{\partial \dot{y}^2} = -\frac{\lambda^2}{(2\pi)^2} \frac{\epsilon_{xx}}{\epsilon_{yy}^2} k_x^3 \,. \tag{S.78}$$

Assuming propagating waves, for which $k_x > 0$, and noticing that all quantities above are non-negative, we obtain

$$\frac{\partial^2 \mathcal{L}}{\partial \dot{y}^2} < 0 \,, \tag{S.79}$$

meaning that, in contrast with isotropic media, geodesics in hyperbolic media are associated with maximum Action.

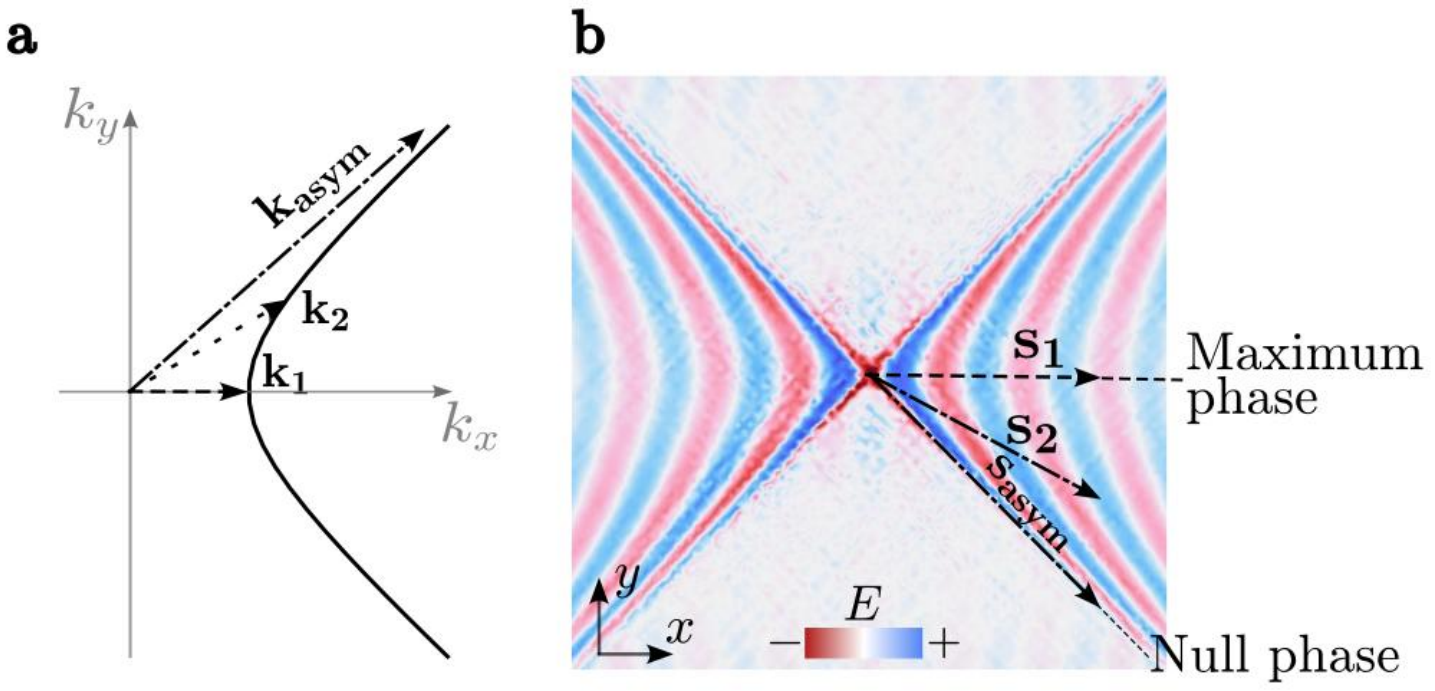

**Figure S3: Inverse Fermat Principle in Hyperbolic Dispersion.** (a) Owing to hyperbolic dispersion, rays $s_1$, $s_2$, and $s_{\mathrm{asym}}$, associated with momenta $k_1$, $k_2$, and $k_{\mathrm{asym}}$, accumulate different phases. By the inverse Fermat principle, for equal horizontal path lengths the on-axis ray $s_1$ acquires the largest phase, while off-axis rays accumulate progressively smaller phases, vanishing in the asymptotic ray $s_{\mathrm{asym}}$, which propagates along the light-cone direction and exhibits frozen phase.

### *2.4 Frozen phase and connection with lens curvature*

The expression in Eq. (S.79) clarifies why a lens formed between hyperbolic media must have a curvature opposite to that of an isotropic lens. As discussed earlier, phase accumulation is largest along the lens's symmetry axis and progressively decreases for rays propagating at increasingly oblique angles, ultimately reaching its minimum as the ray momentum approaches the asymptotic value $k_x \to \infty$ [Fig. S3a-b]:

$$\left(k_y\right)_{asym} = \sqrt{\frac{\epsilon_{xx}}{\epsilon_{yy}}} k_x \,. \tag{S.80}$$

Using the definition of Lagrangian, we define the asymptotic velocity

$$\dot{y}_{asym} = \sqrt{\frac{\epsilon_{yy}}{\epsilon_{xx}}} \tag{S.81}$$

Then, substituting this specific slope into the Lagrangian:

$$\mathcal{L}(\dot{y}_{asym}) \propto \sqrt{1 - \frac{\epsilon_{xx}}{\epsilon_{yy}} \left( \sqrt{\frac{\epsilon_{yy}}{\epsilon_{xx}}} \right)^2} = \sqrt{1-1} = 0 \,, \tag{S.82}$$

This observation implies that a ray propagating exactly along an asymptotic direction would accumulate zero phase, regardless of the distance traveled. In a wave physics interpretation, this vanishing of optical action along asymptotic directions reflects the wave vector becoming orthogonal to the energy-flow direction for which $\mathbf{k} \cdot \mathbf{r} = 0$. In this geometric interpretation, the asymptotes act as the lowest points of the action, where Eq. (S.66) vanishes, whereas the optical axis corresponds to the point of maximal phase accumulation. Consequently, propagation along the asymptotes does not accumulate phase, while any other direction allows a finite phase to build up, with larger accumulation occurring closer to the optical axis. This behavior implies that rays launched at large angles must recover phase in the second medium; to enable this, the lens interface must bend toward the source so that these rays can increase their accumulated phase [Fig. S2]. In the same way that Snell's law minimize the action in isotropic material, the corresponding laws for refraction and reflection can be shown to maximize the action between two points in hyperbolic media (see also Eqs. (S.125) and (S.126) in Sec 4.3).

An additional insight arises from a relativistic-mechanics interpretation, in which the optical Lagrangian of the hyperbolic medium plays the role of a proper-time element [20]:

$$d\tau = \epsilon_{yy} dt^2 - \epsilon_{xx} dx^2 \tag{S.83}$$

In this picture, the optical axis corresponds to a massive particle at rest, $\dot{y} = 0$, and the ray moves purely along the time direction, maximizing its proper time, i.e., $d\tau = \epsilon_{yy} dt^2 > 0$. At the asymptotic angle ( $\dot{y} = \dot{y}_{asym}$ ), the ray behaves like a photon traveling at the speed of light, where its trajectory lies on the light cone, and its proper time interval vanishes ( $d\tau = 0$ ). No "optical aging" occurs along this path.

### *2.5 Lens with focus at finite distance*

The design above, can be generalized to a lens focusing on any coordinate, rather than infinity, as shown in [Fig. S2b]. Assuming the image focus is located at the finite distance $P\left(x_f{}', y_f{}'\right)$, the phase along the optical path is

$$\phi_{OA} + \phi_{OC} = \tfrac{1}{2} + T\sqrt{\left(x_f{}' - \tfrac{1}{2}\right)\left(y_f{}' - \tfrac{1}{2}\right)} \tag{S.84}$$

while other paths contribute to

$$\phi_{OB} + \phi_{BC} = \sqrt{x'y'} + T\sqrt{\left(x_f{}' - x'\right)\left(y_f{}' - y'\right)} \tag{S.85}$$

Assuming the focal point is on the optical path, we have $y_f{}' = x_f{}'$, and the lens equation is

$$y_c{}' = \frac{A(x') + 2T\left[1 + T\left(2x_f{}' - 1\right)\right]B(x')}{4\left[T^2\left(x_f{}' - x'\right) + x'\right]^2}, \tag{S.86}$$

where

$$\begin{aligned} A(x') &= x' - \left\{T^2\left[\begin{matrix} T^2\left(x_f{}' - x'\right)\left(4x_f{}'(x' - 1) + 1\right) + 2T\left(x_f{}' - x'\right)\left(2x_f{}' - 1\right) \\ -8\left(x_f{}'\right)^2 x' + x_f{}'(2x' + 1)^2 - 2x' \end{matrix}\right] + 2Tx'\left(1 - 2x_f{}'\right)\right\}, \\ B(x') &= \sqrt{x'(2T - 1)\left(x_f{}' - x'\right)\left[1 - T - 4Tx_f{}'(x' - 1) - 4x_f{}'x'\right]}. \end{aligned} \tag{S.87}$$

After applying the inverse transformation, the resulting equation can be solved numerically.

## 3. Hyperbolic lens' welding part

### *3.1 Minkowski space geometric approach*

In this section, we derive the equation for the back portion of the hyperbolic lens, termed the *welding region*, designed to minimize diffraction at the line-of-sight point (S.62). To determine the shape of this region, we assume that rays striking the interface beyond this point propagate smoothly along it as lateral waves, eventually reaching the line-of-sight without reflection from a cusp [Fig. S2a]. Specifically, we impose that any ray hitting the welding region accumulates the same total phase as a ray directly reaching any other point along the same line.

To this end, we calculate the phase accumulated by a ray along the curved path using the integral form of Eq. (S.63), which is a generalization of Fermat's principle, as shown previously. We consider the infinitesimal path along the interface $\gamma_w'$ [Fig. S4, blue curved line] expressed as

$$ds = \sqrt{2dx'dy'} = \sqrt{2\frac{dy'}{dx'}}dx' \tag{S.88}$$

Then, using our definition of phase accumulation, the welding interface is the surface that satisfies:

$$\phi_{OD}|_{\gamma_w'} = \frac{2\pi}{\lambda}\sqrt{\frac{\epsilon_{y,2}}{2}}\int_{\gamma_w'} ds = \frac{2\pi}{\lambda}\sqrt{\epsilon_{y,2}}\int_0^{x'}\sqrt{2\frac{dy'(u)}{du}}\,du. \tag{S.89}$$

Were $u$ is the integration variable in the $x$ direction. The path $y'(u)$ is unknown, however we assume the polynomial ansatz

$$y_w'(x) = Cx'^{a}, \qquad 0 < x' < x_L', \tag{S.90}$$

where $C$ is a constant, $x'$ a first order polynomial and $a$ the order of the polynomial. To find the polynomial order $a$, we equate Eq. (S.89) with the phase accumulated along the straight path OD [Fig. S4, red arrow] that is inside the lens and is given by:

$$\phi_{OD}|_{straight} = \frac{2\pi}{\lambda}\sqrt{\frac{\epsilon_{y,1}}{2}}\sqrt{x'y'}, \tag{S.91}$$

Then, imposing

$$\phi_{OD}|_{\gamma_w'} = \phi_{OD}|_{straight} \tag{S.92}$$

and inserting the ansatz, we obtain the expression for the coefficient $a$.

$$a = \left(T - \sqrt{T^2 - 1}\right)^2 \tag{S.93}$$

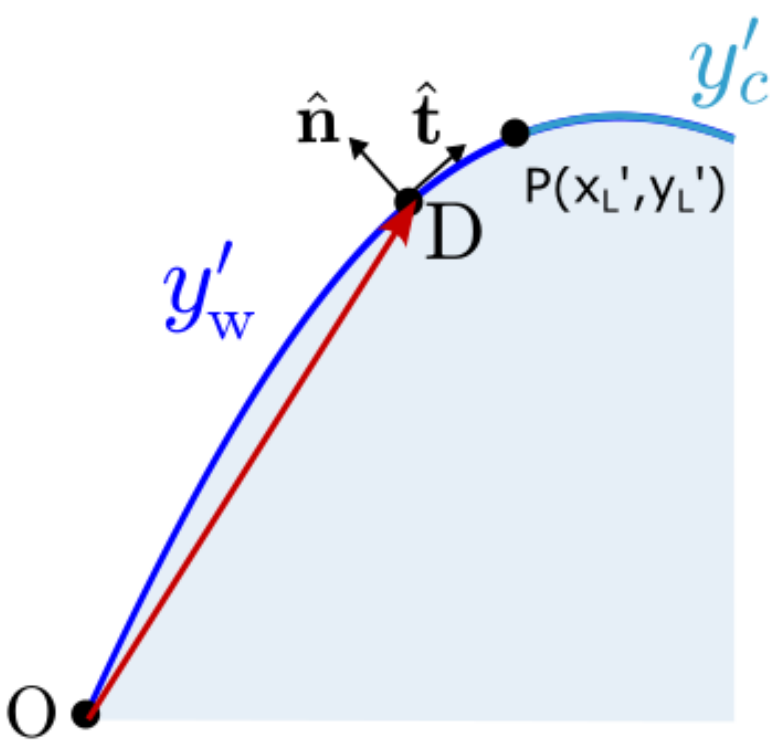

**Figure S4: Welding interface for hyperbolic lens.** The welding region is designed assuming that the refracted ray power is locally parallel to the interface, ensuring refracted rays smoothly exit parallel to the lens interface, minimizing losses and diffraction effects.

Then, we set the coefficient $C$ to smoothly connect the welding part $y_w(x')$ with the collimating part $y_c'(x')$ (found in previous paragraphs), that is equivalent to impose the continuity at the point $(x_L', y_L')$:

$$y_w'(x_L') = y_c'(x_L'), \tag{S.94}$$

finding:

$$C = \frac{x_L^{1-a}}{a} \tag{S.95}$$

The resulting curve is reported as a solid blue line in Fig S2a-b and Fig. S4. In Fig. S5, we examine the impact of the welding region on lens performance. Figure S5b shows the lens without the welding region: in this configuration, power funneling into the aperture is less efficient, and diffraction and reflection losses occur from the corners. By contrast, Fig. S5a demonstrates that including the welding region substantially enhances power collection into the aperture and minimizing diffraction from the edges.

To further justify the construction of the welding region, we detail why rays reaching this portion of the interface must propagate tangentially to it. The phase accumulated by any ray along a path $\gamma_w'$ is defined through the Action integral

$$S_\gamma = \int_\gamma \mathbf{q}\cdot d\mathbf{r}' = \int_\gamma \mathbf{q}\cdot\hat{\mathbf{t}}\,ds' \tag{S.96}$$

We prescribed that, along the straight path, the Action is fixed and given by

$$S = \frac{2\pi}{\lambda}\sqrt{\frac{\epsilon_{y,1}}{2}}\sqrt{x'y'} \tag{S.97}$$

Because this action must come entirely from the motion along the interface, the integrand of Eq. (S.96) must equal the full magnitude of $\mathbf{q}$. In fact, assuming that the wave vector can be written in terms of parallel and normal component to the interface, i.e.,

$$\mathbf{q} = q_{\parallel}\hat{\mathbf{t}} + q_{\perp}\hat{\mathbf{n}}\,, \tag{S.98}$$

then we would have the infinitesimal Action increment only due to the tangential component of the wave vector:

$$dS = \mathbf{q}\cdot d\mathbf{r}' = q_{\parallel}ds\,. \tag{S.99}$$

However, if $q'_{\perp} \neq 0$, it would not contribute to the total Action, meaning that a phase gradient would exist normal to the interface. Because in the Minkowski geometry framework ray and wavefront are orthogonal, this phase gradient would correspond to a wave trying to escape the interface along its normal, which could not happen for the lateral wave we assumed, therefore, the conclusion that

$$\mathbf{q} \parallel \hat{\mathbf{t}}\,. \tag{S.100}$$

In practice, the welding region does not support a genuine guided lateral mode. Instead, the effect arises because the action in each hyperbolic medium is minimized only by straight trajectories in the corresponding Minkowski geometry. When a ray reaches the welding interface at angles beyond the collimation point, any refracted path that bends back into the lens would increase the action and is therefore forbidden. As a result, the only admissible refracted direction is the one tangent to the interface, which preserves the straight-line character of the trajectory in the local Minkowski frame. Thus, rather than forming a bound surface wave, each incident ray is locally refracted into a direction parallel to the boundary, suppressing reflection or scattering at the welding point.

### *3.2 Euclidean space-based approach*

We consider the interface separating an inner medium from an outer medium in Fig. S3, both characterized by arbitrary hyperbolic dispersion relations. We seek the trajectory of the curve $\gamma_w$ connecting the aperture limit $L$ to the source. We postulate that the physical interface is the locus of points where the accumulated optical action (phase) of the incident ray along the straight path OD matches the accumulated action of the ray traversing the external surface. Then:

$$\int_O^D \mathbf{k}_{inc}\cdot\hat{\mathbf{t}}dl = \int_O^D \mathbf{k}_{tra}\cdot\hat{\mathbf{t}}dl \tag{S.101}$$

where $\mathbf{k}_{inc}$ is the incident wave vector, $\mathbf{k}_{tra}$ is the wave vector of the lateral wave travelling along the curved interface, and $\hat{\mathbf{t}}$ the tangent to the curve. Note that $\mathbf{k}_{tra}$ is not necessarily tangent to the interface. Since this conservation must hold for any point $O$ along the trajectory, to ensure phase continuity, the integrand must be continuous and differentiable. Hence, differentiating both sides by the arc length parameter *dl*, we have:

$$\mathbf{k}_{inc}\cdot\hat{\mathbf{t}} = \mathbf{k}_{tra}\cdot\hat{\mathbf{t}} \tag{S.102}$$

Assuming the curve is parametrized as $\gamma_w = (x, y_w)$, we find the tangent vector as:

$$\hat{\mathbf{t}} = \frac{\left(1, y_w'\right)}{\sqrt{1+\left(y_w'\right)^2}} \tag{S.103}$$

where we use the contracted notation $y_w' = \dfrac{dy_w}{dx}$. Hence, from Eq. (S.102):

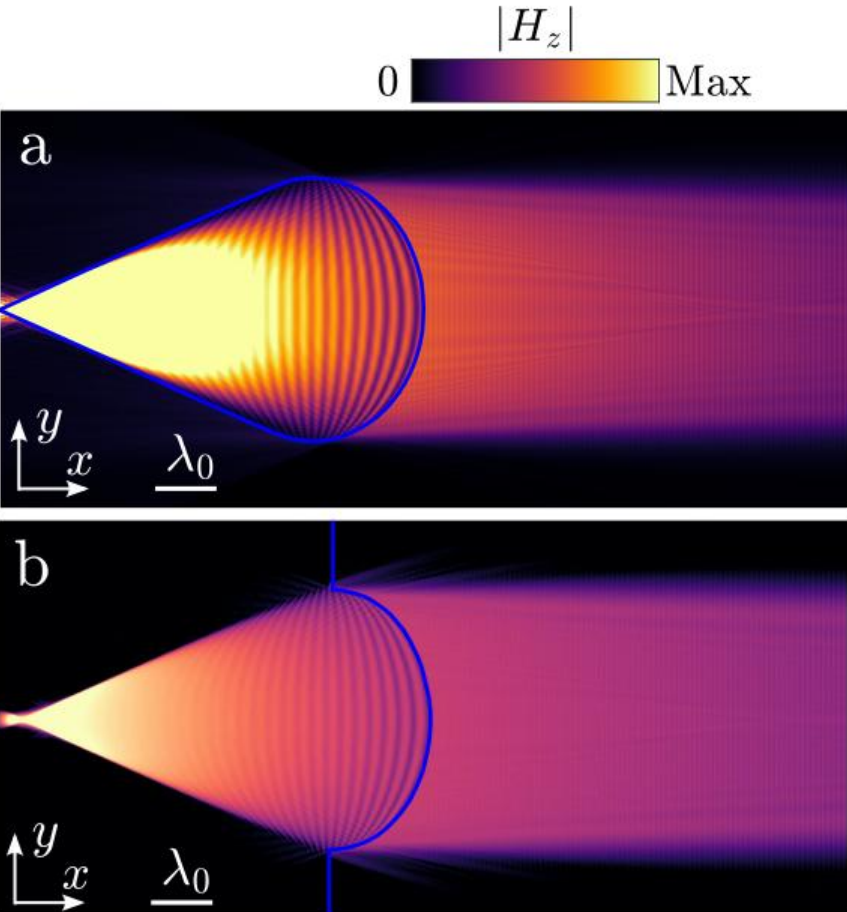

**Figure S5: Full-wave simulations comparing lens configurations.** (a) Hyperbolic lens composed of collimating and welding regions between two 2D non-dispersive hyperbolic media with permittivity tensors $\hat{\epsilon}_1 = \mathrm{diag}([-5,1])$ and $\hat{\epsilon}_2 = \mathrm{diag}([-40,8])$. The welding region significantly reduces diffraction at the edges. (b) Lens composed of the same collimating region, but with a vertical boundary introducing sharp corners. This geometry leads to strong diffraction and stray fields.

$$k_{inc,x} + y_w{}' k_{inc,y} = k_{tra,x}(y_w{}') + y_w{}' k_{tra,y}(y_w{}') \tag{S.104}$$

that gives:

$$y_w{}' = \frac{k_{tra,x}(y_w{}') - k_{inc,x}}{k_{inc,y} - k_{tra,y}(y_w{}')} \tag{S.105}$$

This is an implicit equation for the slope of the interface, $y_w{}'$, as a function of the incident wave vector $\mathbf{k}_{inc}$. Let the power flow of the lateral wave travelling along $\gamma_w'$ be $\mathbf{s}_{tra}$, and let the normal to the interface be:

$$\hat{\mathbf{n}} = \frac{\left(-y_w{}',1\right)}{\sqrt{1+\left(y_w{}'\right)^2}} \tag{S.106}$$

We then require the power-flow direction to be tangent to the interface, effectively enforcing the condition for total internal reflection, that is:

$$\mathbf{s}_{tra} \cdot \hat{\mathbf{n}} = -s_{tra,x} \frac{\partial y_w'}{\partial x'} + s_{tra,y} = 0 \tag{S.107}$$

Then, inverting:

$$\frac{\partial y_w'}{\partial x'} = \frac{s_{tra,y}}{s_{tra,x}} = \tan \beta_{tra} \tag{S.108}$$

Therefore, combining Eq. (S.108) with Eq. (S.105), we obtain:

$$y_w{}' = \frac{k_{tra,x}\left(\tan^{-1} y_w{}'\right) - k_{inc,x}}{k_{inc,y} - k_{tra,y}\left(\tan^{-1} y_w{}'\right)} \tag{S.109}$$

After determining the interface slope $y_w{}'$ solving Eq. (S.109) numerically for each corresponding each injection angle $\beta_i = y'/x'$, we integrate $y_w{}'$ and solve for $x$ numerically, using iterative methods. For example, a possible implementation uses the following fixed-point iteration:

$$x_{new}(i) = x_{old}(i) - \frac{x_{old}(i) - \frac{1}{y'_W(i)}\left(y_L + \int_0^{x_{old}(i)} y'_W(i)dx\right)}{1-y'_W(i)} \tag{S.110}$$

The complexity of this procedure highlights the importance of the simplified approach proposed above where the simple analytical solution in (S.90) can be used. By leveraging Minkowski geometry, our method greatly streamlines the lens design, circumventing the need for iterative numerical solutions.

## 4. Hyperbolic lens' numerical aperture

### *4.1 Maximal collection angle*

In this section we define the numerical aperture of the hyperbolic lens derived previously. Using (S.62) we define the maximal acceptance angle of the lens as

$$\beta_L = sin^{-1}\left(\frac{y_L}{\sqrt{x_L^2+y_L^2}}\right) = sin^{-1}\left(\sqrt{\frac{T^2-1}{T^2(m-1)-1}}\right), \tag{S.111}$$

as shown in Fig. S6a. The expression above was derived using purely geometric considerations. However, an equivalent result can be obtained by considering wave propagation.

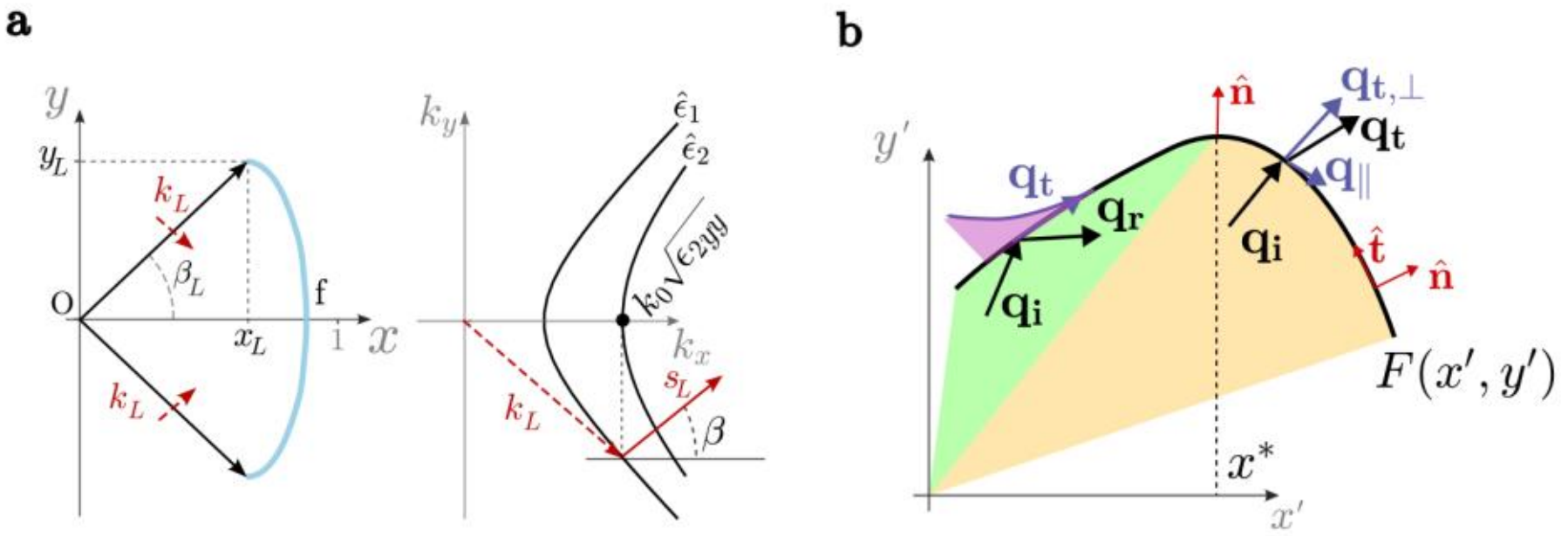

**Figure S6: Numerical-Aperture Limits and Refraction in Light-Cone Coordinates.** (a) (Left) A hyperbolic collimator accepts incident rays within an angular range up to $\beta_L$, which defines the line-of-sight point ($x_L$,$y_L$). (Right) This limiting angle corresponds to the maximum admissible incident momentum $k_L$, matching the vertex of the hyperbolic dispersion outside the lens. The achievable numerical aperture is further constrained by geometry: the local slope of the interface determines whether incident momenta are transmitted or rejected. (b) Refraction at an interface $F(x',y')$ expressed in light-cone coordinates. For space-like normals oriented toward positive $x'$ (yellow sector), incident waves $\mathbf{q}_i$ are refracted to $\mathbf{q}_t$ without the possibility of total internal reflection. Beyond the point where the interface normal, $\hat{\mathbf{n}}$ becomes vertical (green sector), incident waves may undergo negative refraction or total internal reflection.

We observe that the lens interface at the line of sight is tangent and parallel to the $x$-axis [Fig. S6a, left]. Momentum matching requires that the parallel component of the wave vector is conserved across the interface [Fig. S6a, right]. Thus, using the first equation in (S.24), the maximal angle is found solving

$$k_x^{inc}(\beta_L) = k_L = k_0\sqrt{\epsilon_{2,yy}}, \tag{S.112}$$

where $k_L$ is the wave vector of the ray in the medium inside the lens. Eq. (S.112) comes from the fact that the lens contour is parallel to the $x$ axis at the point $(x_L, y_L)$ [Fig. S2a, inset], hence the momentum component to match, is simply the one parallel to this axis, i.e., $k_L$. Inserting Eq. (S.24) we obtain the maximal angle

$$\beta_L = cos^{-1}\left(\sqrt{\frac{\epsilon_{1,xx}\epsilon_{2,yy}}{\epsilon_{2,yy}(\epsilon_{1,xx}+\epsilon_{1,yy})-\epsilon_{1,yy}^2}}\right). \tag{S.113}$$

### *4.2 Geometric numerical aperture*

The geometric numerical aperture is expressed as the maximal normalized transvers momentum, i.e., $k_y$, captured by the interface in the incidence medium. Using the angle in Eq. (S.113), we find:

$$\text{NA} = \frac{k_y^{inc}(\beta_L)}{k_0} = \sqrt{\epsilon_{1,xx}\left(\frac{\epsilon_{2,yy}}{\epsilon_{1,yy}} - 1\right)} = \sqrt{\epsilon_{1,yy}m(T-1)} \tag{S.114}$$

Then for a hyperbolic lens, the numerical aperture is an increasing function of the contrast, and aperture of the iso-frequency contours. By contrast, for lens built between isotropic media, we obtain:

$$\mathrm{NA}_{iso} = \sqrt{\epsilon_1(1-1/T)}, \tag{S.115}$$

which is a limited function of the contrast and incidence medium index only. Eq. (S.115) is obtained assuming that the lens is built between isotropic media, i.e., $\epsilon_{xx} = \epsilon_{yy} = \epsilon > 0$.

### *4.3 Reflection and Refraction at the lens interface*

We perform our analysis in the Minkowski space used to construct the lens of Figs. 2a-b, endowed with metric $\hat{M}'$ in Eq. (S.39), where wave vectors and displacement vectors take the forms given by Eqs. (S.37) and (S.38), and $[\Delta x', \Delta y']^T$, respectively. We remind the reader that this metric space removes dependency on the k-space aperture in the two media. We assume that the interface profile is known and is specified by an implicit equation $F(x', y')$ [Fig. S2a]. Regardless of the choice of metric space, the normal to the interface is naturally represented as a covariant vector, given by

$$\hat{\mathbf{n}} = \nabla_{(x',y')} F(x', y') = \begin{pmatrix} n_{x'} \\ n_{y'} \end{pmatrix}, \tag{S.116}$$

while the tangent to the interface is a contravariant vector, given by

$$\hat{\mathbf{t}} = \begin{pmatrix} dx \\ dy \end{pmatrix} \tag{S.117}$$

To ensure that the tangent and normal vectors are orthogonal with respect to the metric $\hat{M}'$, we exploit the fact that one object is a covector and the other is a vector. Orthogonality is therefore imposed by requiring their contraction (without the use of the metric) to vanish. Explicitly, the tangent vector is determined by enforcing

$$\hat{\mathbf{n}} \cdot \hat{\mathbf{t}} = n_{x'} t_{x'} + n_{y'} t_{y'} = 0 \Rightarrow \hat{\mathbf{t}} = \begin{pmatrix} -n_{y'} \\ n_{x'} \end{pmatrix} \tag{S.118}$$

In Euclidean space, vectors and covectors can be naturally identified via the Euclidean metric, and thus they belong to the same space [39]. As a result, there is no need to distinguish explicitly between covariant and contravariant objects. This distinction becomes essential only in a general (non-Euclidean) metric space. For our lens interface, including the collimating section in Eq. (S.58) and welding section in Eq. (S.90), we have:

$$n_{x'} = \begin{cases} -aC\left(x'\right)^{a-1}, & x' < x'_L \\ T - \sqrt{\dfrac{y'}{x'}}, & x' \geq x'_L \end{cases} \tag{S.119}$$

and

$$n_{x'} = \begin{cases} 1, & x' < x'_L \\ T - \sqrt{\dfrac{x'}{y'}}, & x' \geq x'_L \end{cases} \tag{S.120}$$

We now consider an incident wave vector $\mathbf{q}_i = (q_{i,x'}, q_{i,y'})^T$ associated with a plane wave impinging on the interface. Since the wave vector is defined as the gradient of the wave phase, it is naturally a covariant vector. Consequently, the component of the wave vector parallel to the interface, which must be conserved upon incidence, is obtained by the contraction:

$$q_{\parallel} = \mathbf{q}_i \cdot \hat{\mathbf{t}} = q_{i,y'} n_{x'} - q_{i,x'} n_{y'} \tag{S.121}$$

On the other hand, since both the wave vector and the interface normal are covariant vectors, their normal component cannot be obtained directly by contraction.

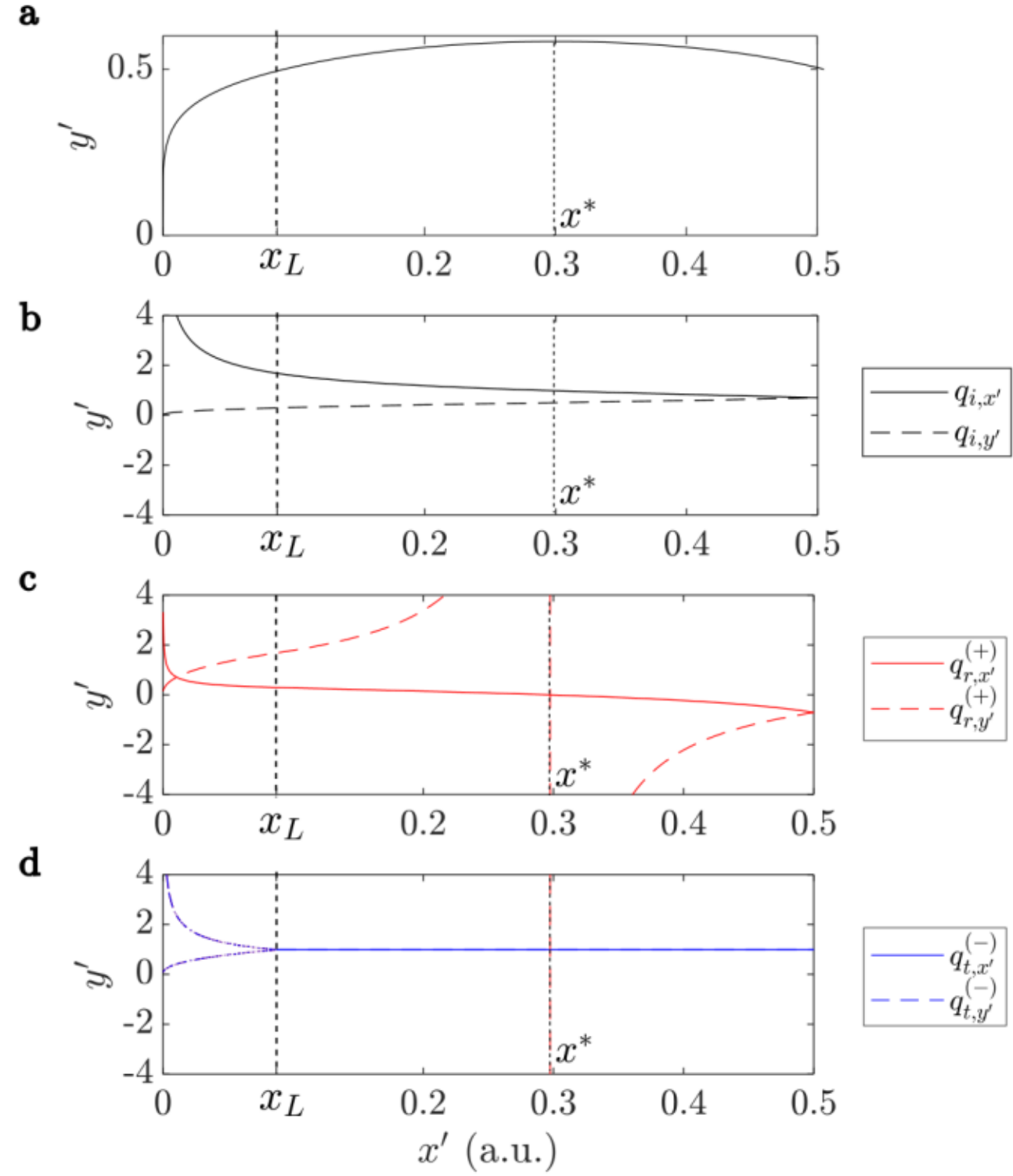

**Figure S7: Reflected and transmitted wave vectors from hyperbolic lens interface.** (a) The lens interface $F(x',y')$ in Eq. (S.58) and Eq. (S.90) is shown for $0 < x' < 1/2$, corresponding to the upper interface. An incident wave with wave-vector components $q_x$ (solid line) and $q_y$ (dashed line) undergoes (c) negative reflection and (d) refraction in the region $x^* < x' < 1/2$, while it is positively reflected for $0 < x' < x^*$. The refracted wave vector is parallel to the $x' = y'$ line for $x_L < x' < 1/2$ due to the collimating interface and becomes tangential for $0 < x' < x_L$. The magnitude of the reflected wave vector diverges at $x^*$, since the only matched component corresponds to the asymptotic momentum of the hyperbolic dispersion branch.

To compute the normal component of the wave vector, the normal must first be mapped to its corresponding contravariant vector $\hat{M}'^{-1}\hat{\mathbf{n}}$. Notice that with our convenient choice of light cone coordinates, the metric satisfies $\hat{M}' = \hat{M}'^{-1}$. Then the normal component is obtained by taking the metric-induced inner product, namely:

$$q_\perp = \mathbf{q}_i \cdot \hat{M}'\hat{\mathbf{n}} = q_{i,x'}n_{y'} + q_{i,y'}n_{x'} \tag{S.122}$$

Using Eq. (S.121) and Eq. (S.43), we obtain the quadratic equation

$$2n_{x'}q_{x'}{}^2 - 2q_\parallel q_{x'} - k_0\epsilon_{yy}n_{y'} = 0 \tag{S.123}$$

Then, for the reflected wave we obtain

$$q_{r,x'}^{(\pm)} = \frac{q_\parallel \pm \sqrt{q_\parallel{}^2 + 2n_{x'}n_{y'}k_0^2\epsilon_{1,yy}}}{2n_{x'}}, \qquad q_{r,y'}^{(\pm)} = \frac{k_0^2\epsilon_{1,yy}}{2q_{r,x'}^{(\pm)}}, \tag{S.124}$$

while for the transmitted waves we obtain

$$q_{t,x'}^{(\pm)} = \frac{q_\parallel \pm \sqrt{q_\parallel{}^2 + 2n_{x'}n_{y'}k_0^2\epsilon_{2,yy}}}{2n_{x'}}, \qquad q_{t,y'}^{(\pm)} = \frac{k_0^2\epsilon_{2,yy}}{2q_{t,x'}^{(\pm)}}. \tag{S.125}$$

Note that the parameter $n_{x'}n_{y'}$ defines two distinct physical regimes. If $n_{x'}n_{y'} > 0$, the normal to the interface is *space-like* pointing to positive $x'$ and $y'$. This means that the interface is *flat* enough that waves can pass through it normally, i.e., the ray vector $\hat{M}'\mathbf{q}_i$ can be parallel to $\hat{\mathbf{n}}$ [Fig. 6b, yellow sector]. Then, since the discriminant in Eqs. (S.124) and (S.125) is non-negative, waves are negatively refracted and no total internal reflection can occur. On the other hand, if $n_{x'}n_{y'} < 0$, the normal vector points toward negative *x'*, and it is *time-like*, so the interface is *steep* enough that no propagating ray can hit it normally, i.e., $\hat{M}'\mathbf{q}_i$ is never parallel to $\hat{\mathbf{n}}$ [Fig. 6b, green sector]. In this regime, the determinant can be negative indicating that both refracted propagating and evanescent waves are supported when $q_{\|}^2 \geq -2n_{x'}n_{y'}k_0^2\epsilon_{\{1,2\},yy}$ or $q_{\|}^2 < -2n_{x'}n_{y'}k_0^2\epsilon_{\{1,2\},yy}$, respectively. In the former case, waves undergo positive refraction. In the latter case, total internal reflection occurs, preventing wave propagation into the second medium and giving rise to ghost surface waves [40], which emerge when an interface is coupled to a hyperbolic medium whose optical axis is neither parallel nor orthogonal to the boundary. The turning point for $n_{x'}n_{y'} = 0$, given by

$$x^* = \frac{1}{T(T+1)}, \qquad y^* = \frac{T}{T+1}, \tag{S.126}$$

corresponds to the interface location where incident rays transition from negative to positive refraction [Fig. S7a]. At this point, although the incident momentum remains finite in magnitude [Fig. S7b], the momenta of the reflected and refracted rays diverge, approaching the asymptotes of the hyperbolic dispersion [Figs. S7c-d]. This coordinate also marks the onset of secondary waves which, after being reflected along the interface toward the front of the lens, generate additional contributions that lead to reflection losses at the output, as illustrated in Figs. S8b,e. These losses can be mitigated by increasing the material contrast $T$, which shifts the critical coordinate $x^*$ toward the line-of-sight point $x'_L$. By inspection of Eqs. (S.60) and (S.126), this condition is satisfied only for $T \to \infty$. This behavior is illustrated in Fig. S8e, where increasing the index contrast results in reduced interference in the refracted wave pattern.

To derive the corresponding Fresnel-type reflection coefficient, we make use of Eq. (S.50) and note that the electric field is a covariant vector [41]. Accordingly, we define the component of the electric field tangent to the interface as the contraction

$$E_{tan} = \mathbf{E}\cdot\hat{\mathbf{t}} = \frac{Z_{eff}}{2k_0} H_z \left( q_{x'}n_{y'} + q_{y'}n_{x'} \right) = \frac{Z_{eff}}{2k_0} H_z q_{\perp}. \tag{S.127}$$

Let the incident, reflected, and transmitted magnetic fields be $H_i$, $rH_i$ and $tH_i$, respectively. Enforcing continuity of both the tangential magnetic and electric fields at the interface yields the two equations:

$$\begin{aligned} &\text{Continuity of } H_z: \quad 1 + r = t, \\ &\text{Continuity of } E_{tan}: \quad Z_{eff,1}\left( q_{i,\perp} + r q_{r,\perp} \right) = Z_{eff,2} q_{t,\perp} t. \end{aligned} \tag{S.128}$$

Then, substituting $1 + r = t$ and rearranging we obtain

$$r = \frac{Z_{eff,2}q_{t,\perp} - Z_{eff,1}q_{i,\perp}}{Z_{eff,1}q_{r,\perp} - Z_{eff,2}q_{t,\perp}}. \tag{S.129}$$

where $Z_{eff} = Z_0 \big/ \sqrt{\epsilon_{xx}\epsilon_{yy}}$. Using $\epsilon_{1,xx}T^2 = \epsilon_{2,xx}$ we obtain:

$$r = \frac{q_{t,\perp} - T^2 q_{i,\perp}}{T^2 q_{r,\perp} - q_{t,\perp}}. \tag{S.130}$$

The equation above generalizes the TM reflection coefficient in case of hyperbolic media. Transmission is found as

$$t = 1 + r = \frac{T^2\left( q_{r,\perp} - q_{i,\perp} \right)}{T^2 q_{r,\perp} - q_{t,\perp}}. \tag{S.131}$$

To evaluate the reflected and transmitted power, we consider the normal component of the Poynting vector, which, being proportional to the product of the tangential electric and magnetic fields, is given by

$$S_{\perp}=\frac{1}{2}\mathrm{Re}\left(E_{tan}H_z^*\right)=\mathrm{Re}\left(Z_{eff}q_{\perp}\left|H_z\right|^2\right). \tag{S.132}$$

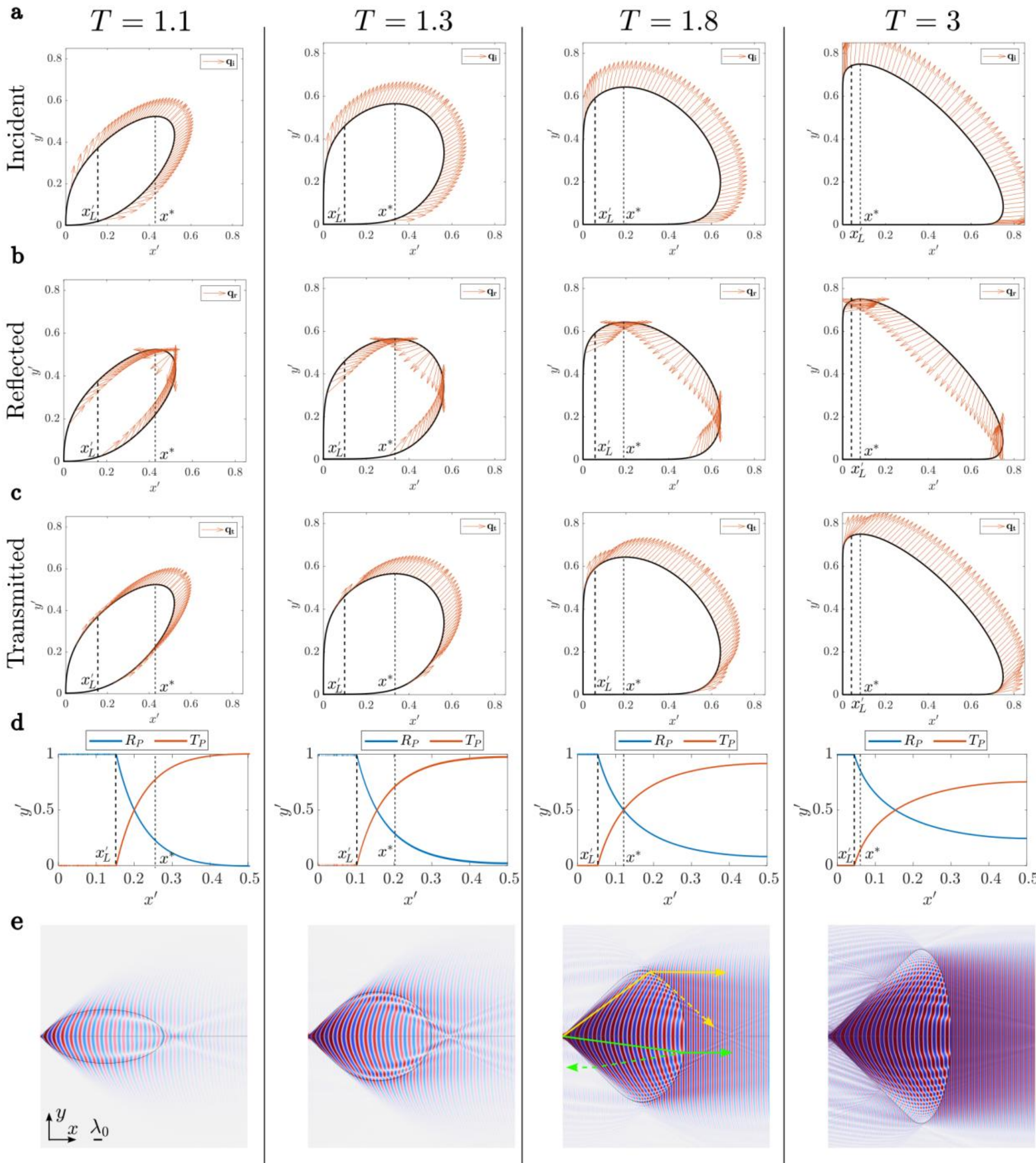

**Figure S8: Effect of Index Contrast on Ray Collimation and Reflection Loss.** Ray dynamics and power balance as a function of index contrast $T$=(1.1,1.3,1.8,3). (a) Incident rays launched from the origin (0,0). (b) Corresponding reflected rays. (c) Corresponding refracted rays. (d) Reflected (red) and transmitted (blue) power, $R_P$ and $T_P$, respectively, showing that increasing index contrast enhances collimation but also increases reflection losses at the front interface of the lens. The line-of-sight coordinate $x'_L$ marks the onset of total internal reflection and shifts toward zero as $T$ increases, thereby maximizing the number of collimated rays while reducing overall transmission efficiency. (e) Full-wave simulations in ($x$,$y$) coordinates ($m$=1), demonstrating enhanced collimation with increasing contrast. The critical coordinate $x^*$ denotes the transition from negative to positive reflection. For $x^*<x<1/2$, rays incident on the interface (green) undergo negative reflection (green dashed lines), whereas for $0<x<x^*$, rays striking the collimating interface (yellow) experience internal reflection (yellow dashed lines), generating secondary refracted waves. As the contrast increases, $x^*$ approaches the line-of-sight point, suppressing secondary-wave-induced reflection losses.

Hence, using $H_r = rH_i$, power reflection is

$$R_P=\left|\frac{S_{r,\perp}}{S_{i,\perp}}\right|=\left|r\right|^2, \tag{S.133}$$

while power transmission is

$$T_P=\left|\frac{S_{\perp,t}}{S_{\perp,i}}\right|=\left|t\right|^2\frac{1}{T^2}\frac{\mathrm{Re}\left(q_{t,\perp}\right)}{q_{i,\perp}}. \tag{S.134}$$

As shown in Fig. S8e, low index contrast results in small reflection losses for primary waves over most incident angles, at the expense of additional losses arising from secondary rays generated by internal reflection. In this regime, the minimum reflection occurs for paraxial rays. Increasing the contrast $T$ enhances direct reflection while suppressing secondary-wave contributions. Since reflection and numerical aperture both increase monotonically with $T$, no optimal maximum exists. Instead, an effective design strategy is to minimize $x^*$, thereby reducing secondary-ray generation while maintaining high numerical aperture. Note that the refraction and reflection results for the lens converge to the isotropic case when the incident wave vector is along the diagonal. In that case, then incident, refracted and reflected rays are parallel to the diagonal and the power transmission can be shown to be the same as for the isotropic case with reflection coefficient

$$r=\frac{T-1}{T+1} \tag{S.135}$$

for the TM waves.

## 5. Green's function for hyperbolic media

### *5.1 Transformed Helmholtz equation for TM waves in anisotropic media*

In this section, we derive the electromagnetic field generated by a current source embedded in an unbounded hyperbolic medium with permittivity tensor $\bar{\varepsilon}=\mathrm{diag}\left(\epsilon_{xx},\epsilon_{yy},\epsilon_{yy}\right)$. The resulting expression provides the rigorous theoretical basis for the lens transfer function derived in the subsequent section.

Here we propose a method to derive the Green's function of a point source radiating within an unbound, non-dispersive, lossless hyperbolic material. This method is inspired by [42], but our approach is based on transformation optics techniques and is limited to TM-polarized waves.

We consider a line current source emitting transverse magnetic TM -polarized waves propagating in the $(x,y)$ -plane. We consider Maxwell equations

$$\begin{gathered}\nabla\times\boldsymbol{E}=i\omega\mu_0\boldsymbol{H}\\ \nabla\times\boldsymbol{H}=-i\omega\bar{\varepsilon}\boldsymbol{E}+\boldsymbol{J}_p\end{gathered} \tag{S.136}$$

Where the source term, $\boldsymbol{J}_p$, is the line current

$$\boldsymbol{J}_p=-i\omega p_y\delta(\boldsymbol{r}-\boldsymbol{r}_0)\hat{\boldsymbol{y}} \tag{S.137}$$

For TM-polarized wave, we assume that the magnetic field is $\mathbf{H}=(0,0,H_z)$, and the electric field is $\mathbf{E}=(E_x,E_y,0)$ Therefore, Eqs. (S.136) become:

$$\begin{gathered}\frac{\partial E_y}{\partial x}-\frac{\partial E_x}{\partial y}=i\omega\mu_0H_z\\ \frac{\partial H_z}{\partial y}=-i\omega\epsilon_{xx}E_x\\ -\frac{\partial H_z}{\partial x}=-i\omega\epsilon_0\epsilon_{yy}E_y+J_y\end{gathered} \tag{S.138}$$

From which we obtain the electric field components:

$$E_x = \frac{1}{-i\omega\epsilon_0\epsilon_{xx}}\frac{\partial H_z}{\partial y}$$
$$E_y = \frac{1}{i\omega\epsilon_0\epsilon_{yy}}\left(\frac{\partial H_z}{\partial x} + J_y\right) \tag{S.139}$$

Combining these in Eq. (S.138), we obtain the inhomogeneous Helmholtz equation:

$$\frac{1}{\epsilon_{yy}}\frac{\partial^2 H_z}{\partial x^2} + \frac{1}{\epsilon_{xx}}\frac{\partial^2 H_z}{\partial y^2} + \omega^2\mu_0\epsilon_0 H_z = -\frac{1}{\epsilon_{yy}}\frac{\partial J_y}{\partial x} \tag{S.140}$$

Because no analytical solution is available for the equation above, we simplify the problem scaling coordinates as

$$u^2 = \epsilon_{yy}x^2, \quad v^2 = \epsilon_{xx}y^2 \tag{S.141}$$

Hence, inserting in Eq. (S.140), we obtain the following modified form of the Helmholtz equation.

$$\frac{\partial^2 H_z}{\partial u^2} + \frac{\partial^2 H_z}{\partial v^2} + \omega^2\mu_0\epsilon_0 H_z = -\frac{1}{\sqrt{\epsilon_{yy}}}\frac{\partial J_y(x,y)}{\partial u}, \tag{S.142}$$

where the current source term is

$$J_y(u,v) = -i\omega p\sqrt{\epsilon_{yy}\epsilon_{xx}}\,\delta(u-u_0)\delta(v-v_0). \tag{S.143}$$

Note that, because of the coordinate transformation, the Helmholtz equation is equivalent to the one describing waves propagating inside an isotropic medium, with the only difference of a rescaled source term.

Let $k_0 = \omega\sqrt{\mu_0\epsilon_0}$, then we obtain the following wave equation with source term:

$$\frac{\partial^2 H_z}{\partial u^2} + \frac{\partial^2 H_z}{\partial v^2} + k_0^2 H_z = S(u,v), \tag{S.144}$$

where the source is

$$S(u,v) = i\omega p\frac{\sqrt{\epsilon_{xx}}}{\sqrt{\epsilon_{yy}}}\delta'(u-u_0)\delta(v-v_0). \tag{S.145}$$

#### *5.2 Calculating the Green's function in homogenous anisotropic media*

Because Eq. (S.144) incorporates a source term, in order to find the electromagnetic fields in the system, we need to determine its Green's function. Subsequently we calculate the field response generated by this localized source. The Green's function is found solving [41]:

$$(\nabla^2 + k^2)G(u,v;u',v') = \delta(u-u')\delta(v-v'), \tag{S.146}$$

which is a 2D Helmholtz equation, for which the Green's function is:

$$G(r) = -\frac{1}{4i}H_0^{(1)}(k_0\rho), \tag{S.147}$$

where $\rho = \sqrt{(u-u')^2 + (v-v')^2}$ and $H_0^{(1)}(k_0\rho)$ denotes the Hankel function of the first kind and zeroth order. Given (S.147), we can find the response to the source integrating the original differential equation in (S.144), which gives:

$$\begin{aligned} H_z(u,v) &= i\omega p\frac{\sqrt{\epsilon_{xx}}}{\sqrt{\epsilon_{yy}}}\iint G(u,v;u',v')\delta'(u'-u_0)\delta(v'-v_0)\,du'dv' \\ &= i\omega p\frac{\sqrt{\epsilon_{xx}}}{\sqrt{\epsilon_{yy}}}\iint G(u,v;u',v_0)\delta'(u'-u_0)\,du' \end{aligned} \tag{S.148}$$

Integrating by parts to remove the derivative of the delta function, and noticing that the total integral vanishes at the boundaries, we obtain:

$$H_z(u,v) = -i\omega p\frac{\sqrt{\epsilon_{xx}}}{\sqrt{\epsilon_{yy}}}\frac{\partial G(u,v;u',v_0)}{\partial u'}\bigg|_{u'=u_0}. \tag{S.149}$$

Noticing that:

$$\frac{\partial G}{\partial u'} = \frac{\partial G}{\partial r}\frac{\partial r}{\partial u'} = -\frac{u'-u}{r}\frac{\partial G}{\partial r} \tag{S.150}$$

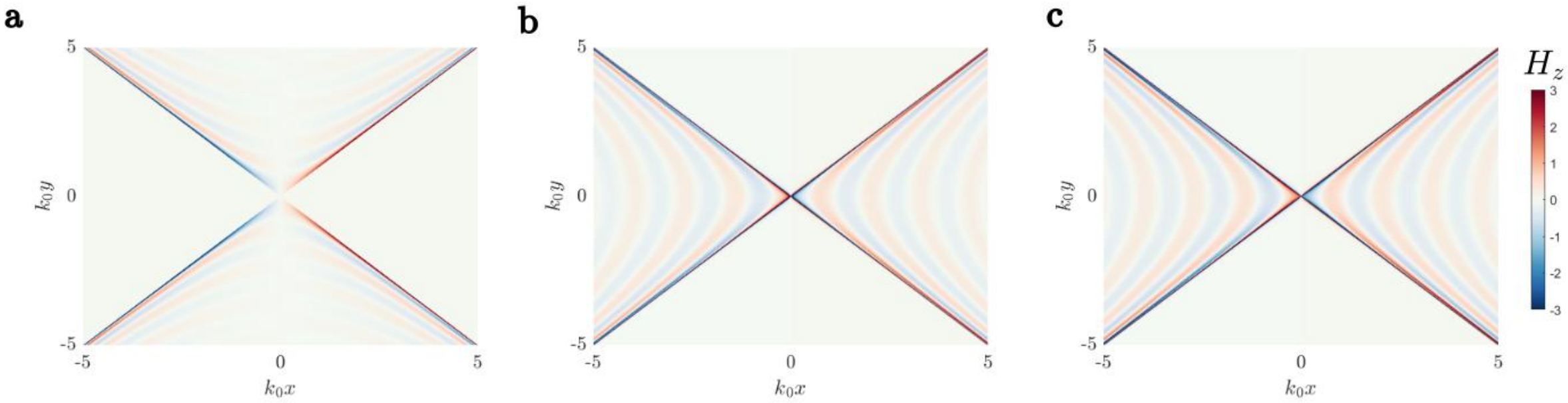

**Figure S9: Validation of the Green's function approximation.** Field distributions of the $H_z$ field (Eq. (S.153)) radiated by point current source inside a hyperbolic medium computed using the exact Green's function (S.147) evaluated for (a) permittivity tensor $\epsilon_1 = (1,-1)$, and (b) permittivity tensor $\epsilon_1 = (-1,1)$. (c) Approximated field distribution (S.157) obtained using (S.154), showing agreement with (b).

Using the property of Hankel function $\frac{d}{ds}H_0^{(1)}(as) = -aH_1^{(1)}(as)$, we obtain:

$$H_z(u,v) = -\frac{\omega k_0 p\sqrt{\epsilon_{xx}}}{4\sqrt{\epsilon_{yy}}}\frac{u-u_0}{\sqrt{(u-u_0)^2+(v-v_0)^2}}H_1^{(1)}\left(k_0\sqrt{(u-u_0)^2+(v-v_0)^2}\right) \tag{S.151}$$

Let $u_0 = 0 = v_0$, then:

$$H_z(u,v) = -\frac{\omega k_0 p\sqrt{\epsilon_{xx}}}{4\sqrt{\epsilon_{yy}}}\frac{u}{\sqrt{u^2+v^2}}H_1^{(1)}\left(k_0\sqrt{u^2+v^2}\right) \tag{S.152}$$

Converting back to familiar $(x, y)$ coordinates:

$$\begin{aligned} H_z(x,y) &= -\frac{\omega k_0 p\sqrt{\epsilon_{xx}}}{4\sqrt{\epsilon_{yy}}}\frac{\sqrt{\epsilon_{yy}}x}{\sqrt{\epsilon_{yy}x^2+\epsilon_{xx}y^2}}H_1^{(1)}\left(k_0\sqrt{\epsilon_{yy}x^2+\epsilon_{xx}y^2}\right) \\ &= H_{z,0}\frac{\sqrt{\epsilon_{yy}}x}{\sqrt{\epsilon_{yy}x^2+\epsilon_{xx}y^2}}H_1^{(1)}\left(k_0\sqrt{\epsilon_{yy}x^2+\epsilon_{xx}y^2}\right) \end{aligned} \tag{S.153}$$

where $H_{z,0} = -\frac{\omega k_0 p\sqrt{\epsilon_{xx}}}{4\sqrt{\epsilon_{yy}}}$ is a constant term. A plot of (S.153) for $(\epsilon_{xx}, \epsilon_{yy}) = (1,-1)$ and $(\epsilon_{xx}, \epsilon_{yy}) = (-1,1)$ is reported in Fig. S9a and Fig. S9b, respectively, showing agreement with Fig. S1b. Note that $\sqrt{\epsilon_{yy}x^2+\epsilon_{xx}y^2}$ is the Minkowski distance defined previously.

Next, to obtain an more easily usable expression of the Green's function (S.147), we approximate assuming that $k_0 r \gg 1$, obtaining:

$$G(r) \simeq \frac{e^{i(k_0 r+\pi/4)}}{\sqrt{8\pi k_0 r}}. \tag{S.154}$$

where $r = \sqrt{\epsilon_{yy}x^2+\epsilon_{xx}y^2}$. Inserting the expression in (S.149), we obtain:

$$H_z(u,v) = -i\omega p\frac{\sqrt{\epsilon_{xx}}}{\sqrt{\epsilon_{yy}}}\frac{\partial G(u,v;u',v_0)}{\partial u'}\bigg|_{u'=u_0}. \tag{S.155}$$

Using:

$$\frac{\partial G(r)}{\partial u'}\bigg|_{u'=u_0} = \frac{\partial G(r)}{\partial r}\frac{\partial r}{\partial u'}\bigg|_{u'=u_0} \simeq -G(r)\left(ik_0 - \frac{1}{2r}\right)\left(-\frac{u-u_0}{r}\right) \tag{S.156}$$

Assuming that $k_0 \gg 1/2r$, we write

$$H_z(u,v) = -\frac{\omega k_0 p\sqrt{\epsilon_{xx}}}{\sqrt{8\pi\epsilon_{yy}}}e^{i\pi/4}\frac{u-u_0}{r^{3/2}}e^{ik_0 r} \tag{S.157}$$

This approximation well matches the behavior of the original expression, as shown in Fig. S9c.

## 6. Lens transfer function

In this section, we derive the transfer function of our hyperbolic lens for two cases: a collimating lens and a converging singlet lens with finite focal distances.

When a plane wave illuminates the lens, the interface acts as a phase mask that reorients the incoming wavevectors toward the focal point [Fig. S10a]. This transformation is accurately modeled by Eq. (S.157). Since we know the field radiated by a point source at the focus, we compute the focusing process through its time-reversed counterpart, interpreting the plane wave after the interface as the converging wavefronts described by Eq. (S.157).

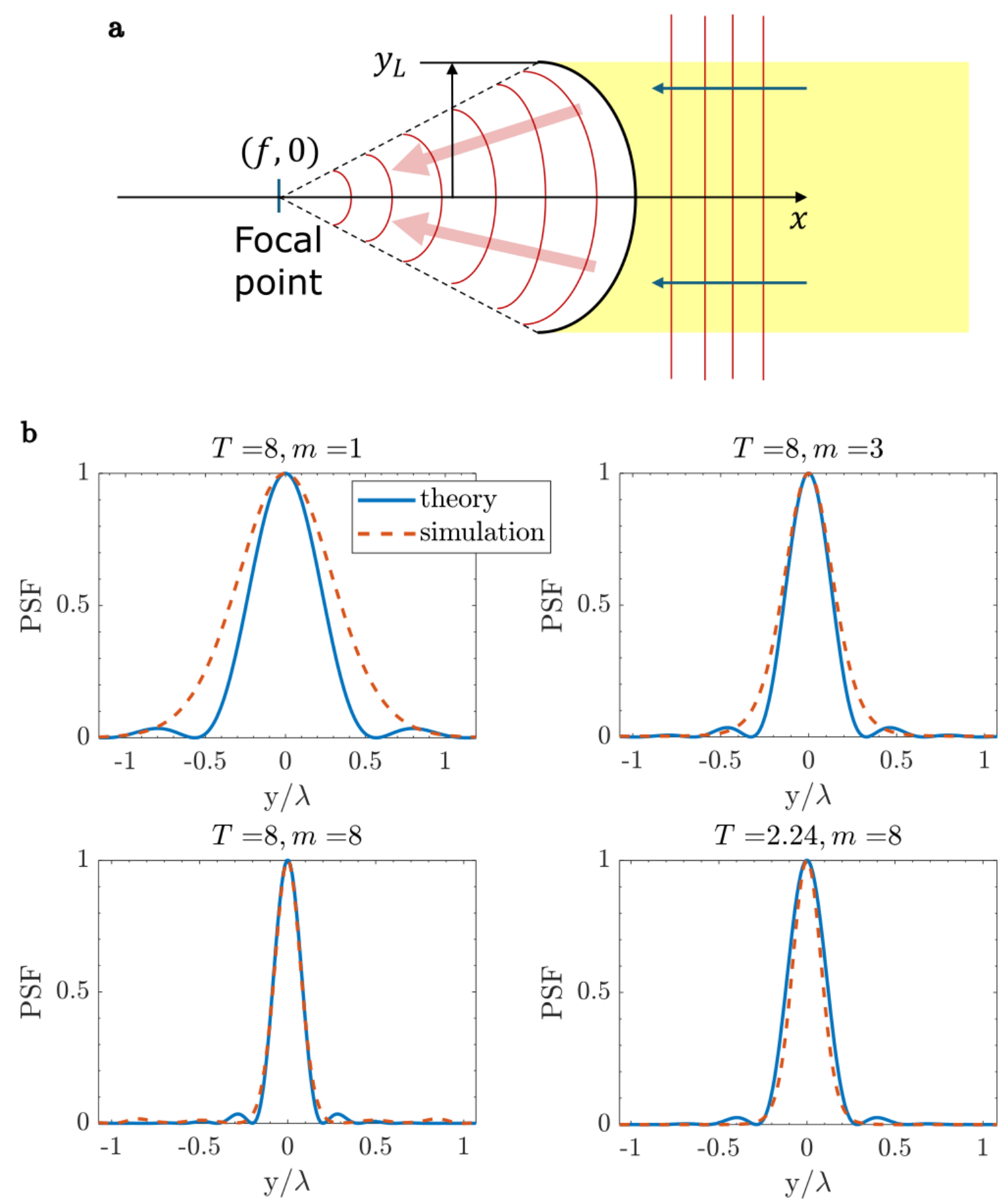

**Figure S10: Hyperbolic lens focusing and resolution analysis.** (a) Pictorial representation of an hyperbolic lens acting as a phase mask for an incoming beam, focusing rays to a focal point located at a distance $f$. The lens is assumed thin ($y_L/f \ll 1$). (b) Calculated lens point spread function (PSF) obtained by solving the generalized Fresnel diffraction integral in hyperbolic geometry (Eq. (S.167)). Results are presented for different values of material contrast $T$ and hyperbolic contour squared aperture $m$. Increasing the values of $T$ and $m$ significantly improves agreement between theoretical (solid blue) and simulated (dashed red) results.

We begin our analysis with a lens that focuses a collimated beam incoming from positive $x$ to a focal point at ($f$,0). Using (S.158), we identify the transfer function of the interface as:

$$T(x,y) = 1_{\odot_a}(y)\,\frac{\sqrt{\epsilon_{yy}}(x-f)e^{-ik_0\sqrt{\epsilon_{yy}(x-f)^2+\epsilon_{xx}y^2}}}{\left[\epsilon_{yy}(x-f)^2+\epsilon_{xx}y^2\right]^{3/2}}, \qquad \text{(S.159)}$$

where

$$1_{\odot_a}(y) = \begin{cases} 1, & y < y_L \\ 0, & y \geq y_L \end{cases} \tag{S.160}$$

is the pupil function of the lens, and $y_L$ is its line of sight (S.62). At the exit pupil of the lens ( $x = 0$ ), we approximate the phase term in (S.159) as:

$$r = \sqrt{\epsilon_{yy} f^2 + \epsilon_{xx} y^2} = \sqrt{\epsilon_{yy}} f \sqrt{1 + \frac{\epsilon_{xx} y^2}{\epsilon_{yy} f^2}} \simeq \sqrt{\epsilon_{yy}} f + \frac{1}{2} \frac{\epsilon_{xx} y^2}{\sqrt{\epsilon_{yy}} f} \tag{S.161}$$

This condition is equivalent to imposing a thin lens condition and assume that the focal distance is larger than the lens aperture. Although this requirement is not always verified, it is valid for large iso-frequency contours aperture and index contrast $T$ between the media. Hence, dropping constant terms we find the transmission function of the lens:

$$\tau(0, y) \simeq 1_{\odot_a}(y)\, exp\left\{-i \frac{1}{2} k_0 \frac{\epsilon_{xx} y^2}{\sqrt{\epsilon_{yy}} f}\right\}, \tag{S.162}$$

which is analogous to a parabolic phase mask imposed by a circular aperture in the Fresnel approximation, but with a rescaling phase coefficient $\epsilon_{xx}/\sqrt{\epsilon_{yy}}$ given by the medium hyperbolicity. Therefore, at the exit pupil, any field distribution $U_0(y)$ is transformed as

$$U_L(y) = U_0(y) 1_{\odot_a}(y) e^{-i \frac{1}{2} k_0 \frac{\epsilon_{xx} y^2}{\sqrt{\epsilon_{yy}} f}} \tag{S.163}$$

Next, we generalize the Fresnel diffraction integral to the case in which we have anisotropy:

$$U(x, y) = \frac{e^{i k_0 \sqrt{\epsilon_{yy}} x}}{i \lambda_0 \sqrt{\epsilon_{yy}} x} \int_{-\infty}^{\infty} U_L(y')\, e^{i \frac{k_0}{2\sqrt{\epsilon_{yy}} x} [\epsilon_{xx} (y - y')^2]} dy'. \tag{S.164}$$

Rearranging the phase term we obtain:

$$U(x, y) \simeq \frac{e^{i k_0 \sqrt{\epsilon_{yy}} x} e^{i \frac{\epsilon_{xx}}{\sqrt{\epsilon_{yy}}} \frac{k_0 y^2}{2\ x}}}{i \lambda_0 \sqrt{\epsilon_{yy}} x} \int_{-y_L}^{y_L} U_0(y') e^{i \frac{\epsilon_{xx}}{\sqrt{\epsilon_{yy}}} \frac{k_0}{2} y'^2 \left(\frac{1}{x} - \frac{1}{f}\right)} e^{-i \frac{\epsilon_{xx}}{\sqrt{\epsilon_{yy}}} k_0 y \frac{y'}{x}} dy'. \tag{S.165}$$

Hence, At the back focal plane ($x = f$), we have:

$$U(f, y) \simeq \frac{e^{i k_0 \sqrt{\epsilon_{yy}} f} e^{i \frac{\epsilon_{xx}}{\sqrt{\epsilon_{yy}}} \frac{k_0 y^2}{2\ f}}}{i \lambda_0 \sqrt{\epsilon_{yy}} f} \int_{-y_L}^{y_L} U_0(y') e^{-i \frac{\epsilon_{xx}}{\sqrt{\epsilon_{yy}}} \frac{k_0 y}{f} y'} dy' \tag{S.166}$$

If a plane wave strikes on the lens, then $U_0(y') = 1$, which yields the solution

$$U(f, y) \simeq \frac{e^{i k_0 \sqrt{\epsilon_{yy}} f} e^{i \frac{\epsilon_{xx}}{\sqrt{\epsilon_{yy}}} \frac{k_0 y^2}{2\ f}}}{i \lambda_0 \sqrt{\epsilon_{yy}} f} 2 y_L \operatorname{sinc}\left(\frac{\epsilon_{xx}}{\sqrt{\epsilon_{yy}}} \frac{k_0 y}{f} y_L\right) \tag{S.167}$$

Notice that the solution of this problem is equivalent to that of a slit that imposes a modified parabolic phase (S.162). Fig. S10b shows how increasing index contrast $T$ and aperture $m$, the expression in (S.167) better approximates the actual point spread function of the lens. We note that this calculation is not valid for circular pupils, however this is not needed in our framework, since we apply our result to planar pupils. To find the resolution of the lens, we remind that we are considering the $\hat{\epsilon}_1$ medium, and calculate the zero of the sinc function:

$$\text{Res} = \frac{\lambda_0}{2} \sqrt{\frac{4(T+1)}{\epsilon_{1,yy} m (T-1)}} \tag{S.168}$$

Then, we find:

$$\text{Res} = 2 \lambda_0 \frac{\sqrt{T+1}}{\text{NA}} \tag{S.169}$$

These results can be generalized for a lens focusing a point source located at $(-f_1, 0)$ to a focus $(f_2, 0)$. To do so, we assume that waves are launched from $(-f_1, 0)$ propagate following (S.157), building the source term $U_0(y')$. The solution of this integral is then found using Wolfram Mathematica, obtaining the field at the focal point $(f_2, 0)$

$$U(y, f_2) \propto exp\left\{i\frac{k_0}{2}\frac{\epsilon_{1,xx}}{\sqrt{\epsilon_{1,yy}}}f_1 + i\frac{k_0}{2}\frac{\epsilon_{2,xx}}{\sqrt{\epsilon_{2,yy}}}\frac{x^2}{f_2} - i\frac{B(x)^2}{4A}\right\}$$
$$\times \frac{1}{2}\sqrt{\frac{\pi}{iA}}\left\{\text{erf}\left[\sqrt{iA}\left(\frac{D}{2} - \frac{B(x)}{2A}\right)\right] + \text{erf}\left[\sqrt{iA}\left(\frac{D}{2} + \frac{B(x)}{2A}\right)\right]\right\} \quad \text{(S.170)}$$

where $A = \frac{k_0}{2}\frac{\epsilon_{1,xx}}{\sqrt{\epsilon_{1,yy}}}\frac{1}{f_1}$ and $B(x) = k_0\left(\frac{\epsilon_{1,xx}}{\sqrt{\epsilon_{1,yy}}} + \frac{\epsilon_{2,xx}}{\sqrt{\epsilon_{2,yy}}}\frac{x}{f_2}\right)$.

Note that in this case, there is no closed form solution for the resolution.

## 7. Loss analysis

We examine the impact of material loss on the proposed model. Resonant media are typically described by a Lorentzian response; in polaritonic systems, this implies a complex permittivity with loss rates on the order of $\sim$1% of the characteristic polariton frequency. In Fig. S11 we consider a representative case with $\epsilon_{yy,1}^{real} = 1$, contrast $T$=2.23 and aperture $m$=1, introducing an imaginary permittivity component $\epsilon_{yy,1}^{imag} = 0.5\%, 1\%, 2\%$ of $\epsilon_{yy}^{real}$ in both materials. These values are realistic for natural materials and metamaterials operated far from resonance, where losses are minimized and nonlocal effects are negligible [12,16-19,36].

Although the analytical framework developed throughout this work assumes lossless media, our full-wave simulations include a small loss term to ensure numerical convergence, yielding results in excellent agreement with theory. Figures S11a,b show that increasing loss primarily leads to attenuation of the field amplitude at the focal point, while simultaneously suppressing secondary reflections. Importantly, the focal position remains fixed at $x$=0 and $y$=0, as predicted by the design strategy, with loss affecting only the overall intensity through propagation damping.

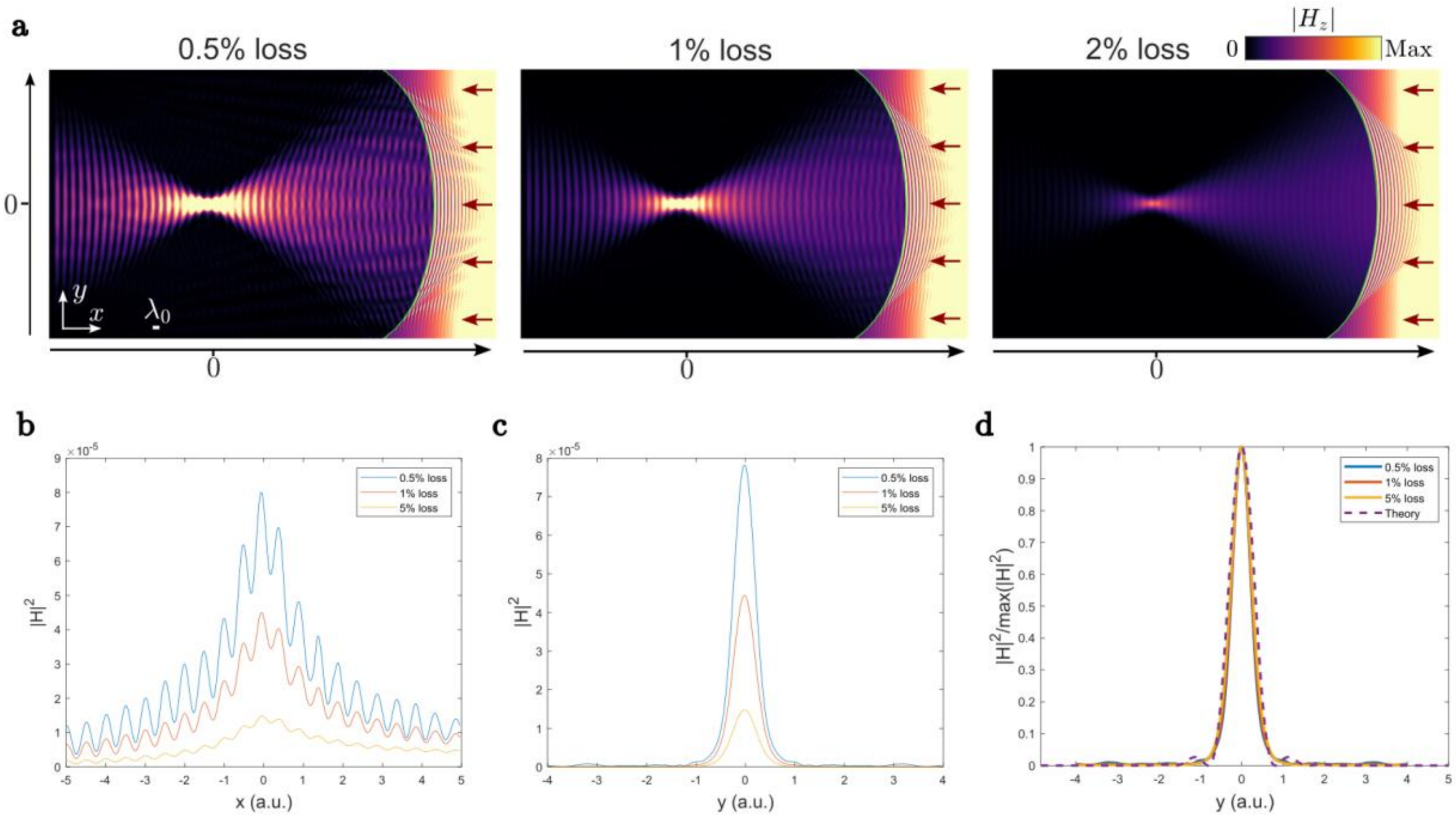

**Figure S11: Robustness of Hyperbolic Lens Focusing in the Presence of Material Loss.** Full-wave validation of lens performance under material loss for a lens with $\varepsilon_{yy,1}$=1, contrast $T$=2.23, and $m$=1. (a) COMSOL-simulated magnetic-field distributions including 0.5%, 1%, and 2% loss in $\varepsilon_{yy,1}$ and $\varepsilon_{yy,2}$. These loss levels are representative of Lorentzian media operated sufficiently far from resonance. (b) Longitudinal field profile along the $x$ axis. (c) Transverse field profile along the $y$ axis, corresponding to the point-spread function (PSF), showing that the focal peak remains fixed in position despite increasing loss. (d) Comparison between the normalized full-wave field on the vertical focal plane (blue, red, and yellow curves) and the theoretical PSF predicted by Eq. (S.167) (purple dashed curve), demonstrating excellent agreement and confirming the robustness of the approach against material losses.

Furthermore, comparison between full-wave simulations and the point-spread function calculated from Eq. (S.167), which neglects loss, reveals excellent agreement, demonstrating the robustness of the method under realistic material losses characteristic of resonant natural and artificial media.

## 8. Polaritonic lens construction

Here we explain the derivation of the lens shape in Fig. 4 of the main text where we apply our theory to the design of a realistic polaritonic nano-lens operating in the mid-infrared, a spectral range central to sensing and imaging applications, where low-loss and ultra-compact architectures are essential [39-41]. While previous studies have explored polaritonic lens concepts, these have either relied on weakly confined modes [33] or employed approximate lens profiles that do not fully capture hyperbolic propagation [36]. Here, we use our Minkowski-based design strategy to realize a compact lens supporting strongly subwavelength focusing, implemented in α-$MoO_3$, a van der Waals (vdW) crystal hosting ultra-low-loss hyperbolic surface phonon polaritons (h-SPhPs) in the mid-IR. This platform is particularly well suited to our approach due to recent advances in nanofabrication that enable precise control of polariton confinement through thickness engineering of α-$MoO_3$ layers [33,36,42,43].

The proposed structure [Fig. 4a] consists of two α-$MoO_3$ regions with different thicknesses: a thicker inner region [Fig. 4a, inset, Region I], supporting less confined modes and acting as an effective low-index medium, and a thinner outer region [Fig. 4a, inset, Region II], supporting strongly confined modes and serving as the high-index hyperbolic medium. Importantly, while thickness variations leave the hyperbolic aperture unchanged, they provide direct control over the effective index, allowing the implementation of the lens profiles prescribed by our theory. To further enhance confinement and minimize the achievable focal spot, the α-$MoO_3$ bilayer is placed on a gold–$SiO_2$ substrate, which supports ultra-confined polaritons [44] and ensures that the guided dispersion closely follows a hyperbolic contour over a broad range of in-plane momenta.

After extracting the frequency- and thickness-dependent contrast, we determine the effective medium parameter by interpolating the guided mode dispersion [35,42]. This procedure yields the effective permittivity tensor $\hat{\epsilon}_{eff}$, from which we find $m_{eff} = 2.4$ and $T_{eff} = 1.73$. Using these parameter, we implement our method and design a lens that focuses a point source located at $x = f_1$ to the focal point $x = f_2$ [Fig. 4b], with $f_1 = -5.65\mu m$ and $f_2 = 2.35\mu m$, and operating at 27.3 THz. The lens aperture spans a width $w = 2.9\mu m$. Full-wave simulations in Fig. 4b show the electric field norm $|E_z|$ while Fig. 4c presents the field distribution at the focal plane (blue) overlaid with the theoretical prediction (dashed red line). The simulated point spread function closely matches the theoretical model, yielding a resolution of $R = \lambda_0 / 42.07 = \lambda_p / 1.91$ ($\lambda_0 = 10.58\mu m$, $\lambda_p = 480nm$), in excellent agreement with our analytical framework.

## 7. References

38. H. Goldstein, C. P. Poole, J. L. Safko, “Classical Mechanics,” 3rd ed, (Pearson Inc., 2001).
39. M. P. d. Carmo, Differential Geometry of Curves and Surfaces (Prentice-Hall, Englewood Cliffs, NJ, 1976).
40. W. Ma, G. Hu, D. Hu, R. Chen, T. Sun, X. Zhang, Q. Dai, Y. Zeng, A. Alù, *et al.*, “Ghost hyperbolic surface polaritons in bulk anisotropic crystals,” *Nature* 596, 362–366 (2021).
41. J. D. Jackson, Classical Electrodynamics, 3rd ed. (Wiley, New York, 1998).

42. A. S. Potemkin, A. N. Poddubny, P. A. Belov, and Y. S. Kivshar, Green function for hyperbolic media, Phys. Rev. A 86, 023848 (2012)